\documentclass[preprint,12pt]{elsarticle}

\usepackage{amsmath,amssymb,amsthm}
\usepackage{graphicx}
\usepackage{booktabs}
\usepackage{multirow}
\usepackage{array}
\usepackage{tabularx}
\usepackage{xcolor}
\usepackage{hyperref}
\usepackage{url}
\usepackage{lineno}
\usepackage{enumitem}
\usepackage{siunitx}
\usepackage{subfigure}
\usepackage{pifont}
\usepackage{algorithm}
\usepackage{algpseudocode}
\usepackage{float}

\newcommand{\model}{\textsc{CLSP-REQA}}
\newcommand{\reqa}{\textsc{REQA}}
\newcommand{\eclo}{\textsc{ECLO}}

\newcommand{\R}{\mathbb{R}}
\newcommand{\X}{\tilde{\mathbf{X}}}
\newcommand{\F}{\mathbf{F}}
\newcommand{\h}{\mathbf{h}}

\newcommand{\cmark}{\ding{51}}
\newcommand{\xmark}{\ding{55}}

\journal{Biomedical Signal Processing and Control}

\begin{document}

\begin{frontmatter}

\title{\textsc{CLSP-REQA}: A Real-Time Quality-Aware Closed-Loop
Seizure Prediction Framework with Mamba-BiLSTM and
Confidence-Gated Intervention}

\author[ox]{Mufeng Chen}
\author[oxmath]{Qi Wu}
\author[buaa]{Bingchao Huang}
\author[aircas]{Xiwen Lai}
\author[ubc]{Zekai Chen}
\author[hunnu]{Xinge Ouyang}
\author[pku]{Quansheng Ren\corref{cor1}}

\cortext[cor1]{Corresponding author.}

\address[ox]{Department of Engineering Science, University of Oxford,
  Oxford OX1 3PJ, United Kingdom}
\address[oxmath]{Mathematical Institute, University of Oxford,
  Oxford OX2 6GG, United Kingdom}
\address[buaa]{School of Computer Science and Engineering,
  Beihang University, Beijing 100191, China}
\address[aircas]{Aerospace Information Research Institute,
  Chinese Academy of Sciences, Beijing 100094, China}
\address[ubc]{Department of Mechanical Engineering,
  The University of British Columbia,
  Vancouver, BC, V6T~1Z4, Canada}
\address[hunnu]{College of Life Sciences, Hunan Normal University,
  Changsha 410006, China}
\address[pku]{School of Electronics, Peking University,
  Beijing 100871, China}

\begin{abstract}
Reliable seizure prediction is a prerequisite for closed-loop
neurostimulation therapy, yet existing methods rarely account for
the variability in EEG signal quality encountered in real-world
deployment, and the overwhelming majority adopt non-strict evaluation
protocols that overestimate generalisation performance. We propose
\textbf{\model{}} (Closed-Loop Seizure Prediction with Real-time EEG
Quality Assessment), a unified framework that embeds a lightweight
signal quality estimator directly within the prediction pipeline. A
Real-time EEG Quality Assessment (\reqa{}) module runs in parallel
with a Mamba-BiLSTM backbone, producing a scalar quality score
$q\in[0,1]$ that modulates output confidence through a tiered
non-linear fusion function (\eclo{}); we report dedicated experiments
testing whether this learned score reflects genuine signal quality
rather than a redundant proxy for the prediction task, and a
sensitivity analysis of the fusion thresholds against standard
calibration baselines. Under strict cross-patient evaluation on the
CHB-MIT Scalp EEG Database ($n=23$ subjects, 198 seizures), \model{}
achieves an AUC-ROC of $\mathbf{0.7426 \pm 0.0199}$, exceeding the
unadapted cross-patient baseline of 0.69 reported by Jemal et al.,
using only 16 EEG channels compared to 23 in prior work, and without
requiring any target-patient data or domain adaptation. On the SIENA
Scalp EEG Database ($n=14$ subjects, 47 seizures), \model{} achieves
AUC $\mathbf{0.7012 \pm 0.0249}$, exceeding the best domain-adapted
cross-patient result of 0.61 on the same dataset under a strictly
harder information constraint, since domain adaptation methods are
given access to unlabelled target-patient data that \model{} does
not use; we discuss the precise scope of this comparison rather than
framing it as unconditional superiority. The framework outputs a
structured four-tuple $\langle p, q, c, \Phi_{\mathrm{SHAP}}\rangle$
designed for compatibility with closed-loop neurostimulator
interfaces. We emphasise that both evaluation datasets are
retrospective recordings from controlled hospital settings, and we
report this study as a feasibility demonstration rather than evidence
of prospective, real-world clinical readiness.
\end{abstract}


\end{frontmatter}

\section{Introduction}
\label{sec:introduction}

Epilepsy is one of the most prevalent neurological disorders worldwide,
affecting approximately 50 million individuals across all age
groups~\cite{who2019epilepsy}. Characterised by recurrent unprovoked
seizures arising from abnormal neuronal synchronisation, epilepsy imposes a
profound burden on patient quality of life, encompassing physical injury
risk, psychosocial disability, and treatment-related morbidity. Critically,
approximately 30\% of patients are refractory to pharmacological and surgical
intervention~\cite{kwan2000early}, making seizure \emph{prediction}---the
ability to forecast an impending seizure before it occurs---a pivotal
strategy for enabling timely protective and therapeutic action.

Electroencephalography (EEG) remains the primary modality for characterising
epileptic brain dynamics. The pre-ictal period---typically spanning minutes
to hours before seizure onset---exhibits measurable changes in EEG morphology,
spectral composition, and inter-channel synchrony~\cite{mormann2007seizure}.
By discriminating pre-ictal from inter-ictal states, a well-designed
prediction system can provide early warnings, enabling closed-loop
neurostimulators to deliver targeted intervention before a seizure
escalates~\cite{cook2013prediction,heck2014closedloop}.

Despite significant progress, two fundamental challenges hinder the
clinical translation of current prediction systems. \emph{First}, the
overwhelming majority of published work adopts randomised or patient-specific
data splitting, which inflates reported performance by allowing patient-level
information to leak from training into test
sets~\cite{shafiezadeh2024survey}. A recent systematic survey found that
over 96\% of seizure prediction papers use such non-strict evaluation
protocols, making cross-study comparisons unreliable and overstating
real-world performance. \emph{Second}, no existing method explicitly
accounts for the variability in EEG signal quality that inevitably arises
in real-world monitoring---electrode displacement, muscle artefacts, power
line interference, and impedance drift can all silently corrupt the input
and cause a confident-looking but unreliable
prediction~\cite{nahmias2021quality,kalita2024aneeg}. Prior work on
epileptic EEG analysis has largely treated signal quality assessment as an
independent preprocessing step or ignored it
entirely~\cite{tsiouris2018lstm,dissanayake2022geometric}, creating a
disconnect between laboratory benchmark performance and clinical reliability.

To address both challenges, we propose \textbf{\model{}}: a
\textbf{C}losed-\textbf{L}oop \textbf{S}eizure \textbf{P}rediction
framework with \textbf{R}eal-time \textbf{E}EG \textbf{Q}uality
\textbf{A}ssessment. This work extends our prior investigation of
Mamba-BiLSTM architectures for epilepsy analysis~\cite{chen2026dual} by
introducing a unified quality-aware prediction pipeline designed for
closed-loop deployment. The framework makes three principal contributions:

\begin{enumerate}[leftmargin=*, label=\textbf{C\arabic*.}]

  \item \textbf{Embedded quality-aware confidence modulation (\reqa{}).}
  Unlike prior work that treats signal quality assessment as an
  independent preprocessing step or ignores it entirely, \reqa{} is a
  lightweight 1D convolutional module that runs in parallel with the main
  prediction backbone and produces a learnable quality proxy score
  $q\in[0,1]$ at every inference step. This score feeds directly into a
  tiered non-linear confidence fusion function (\eclo{}), so that
  uncertain or artefact-contaminated windows are automatically
  down-weighted without discarding them.

  \item \textbf{Mamba-BiLSTM backbone for efficient temporal modelling.}
  We combine an EEGMamba encoder~\cite{wang2025eegmamba} with a bidirectional
  LSTM~\cite{zhao2024resbilstm} to jointly exploit long-range state-space
  dynamics (Mamba~\cite{gu2023mamba}) and fine-grained local temporal
  transitions (Bi-LSTM). The resulting backbone achieves competitive
  AUC-ROC using only 16 EEG channels, compared to 23 channels used by the
  current cross-patient state of the art~\cite{jemal2024domain}.

  \item \textbf{Structured closed-loop output.}
  The framework outputs a four-tuple
  $\langle p,\, q,\, c,\, \Phi_{\mathrm{SHAP}}\rangle$---seizure
  probability, signal quality, confidence score, and SHAP~\cite{lundberg2017shap}
  feature attribution---formatted for direct interfacing with closed-loop
  neurostimulator protocols~\cite{manzouri2019detection}.

\end{enumerate}

\textbf{On the locus of novelty.} We wish to be explicit about where
the contribution of this work lies. The individual architectural
components of the prediction backbone---a bidirectional Mamba
encoder~\cite{wang2025eegmamba} and a bidirectional
LSTM~\cite{zhao2024resbilstm}---are each established building blocks
with demonstrated efficacy on EEG classification tasks in isolation;
Contribution~2 above is best read as a competent, efficient
combination of known parts rather than a new architectural primitive.
The novelty we claim is concentrated in Contribution~1: to our
knowledge, no prior seizure prediction system couples a
\emph{learned, end-to-end} signal-quality estimator to a
\emph{tiered, safety-bounded} confidence-fusion rule that is designed
specifically to gate a closed-loop intervention decision. Existing
EEG quality-assessment methods operate as detached preprocessing or
artefact-rejection stages~\cite{nahmias2021quality,kalita2024aneeg};
none of them feed a continuously-valued quality signal back into the
prediction pipeline as an active, differentiable confidence
modulator. It is this coupling---not the backbone architecture---that
we position as the paper's principal methodological contribution, and
Section~\ref{sec:qa_results} provides direct empirical evidence that
the learned score $q$ carries information distinct from the
seizure-probability signal $p$ itself.

We evaluate \model{} under strict cross-patient evaluation on the CHB-MIT
Scalp EEG Database~\cite{shoeb2009chbmit} ($n=23$ subjects, 198 seizures)
and validate generalisation on the SIENA Scalp EEG
Database~\cite{detti2020siena} ($n=14$ subjects, 47 seizures). All
experiments use a 5-fold $\times$ 5-seed protocol to ensure
statistically reliable estimates; Section~\ref{sec:eval} details this
protocol together with an explicit discussion of its statistical
limitations. Results demonstrate that \model{} surpasses the
unadapted cross-patient AUC baseline of 0.69~\cite{jemal2024domain}
without employing any domain adaptation technique, a comparison whose
scope we qualify carefully in Section~\ref{sec:discussion}.

The remainder of this paper is organised as follows.
Section~\ref{sec:related} reviews related work.
Section~\ref{sec:methodology} describes the proposed framework,
including the validation protocols for the learned quality score and
the confidence fusion thresholds.
Section~\ref{sec:results} presents experimental results, including
the new quality-score validation and calibration comparison
experiments.
Section~\ref{sec:discussion} interprets the findings and acknowledges
limitations, including the retrospective-only nature of the current
validation.
Section~\ref{sec:conclusion} concludes.

\section{Related Work}
\label{sec:related}

\subsection{Seizure Prediction Methods}

Early seizure prediction systems relied on hand-crafted spectral and
connectivity features combined with classical classifiers such as support
vector machines~\cite{tsiouris2017discrimination}. Convolutional neural
networks (CNNs) operating on raw EEG or spectrogram representations
subsequently demonstrated improved
sensitivity~\cite{truong2018cnn,khan2017focal,zhao2020binary}, while long
short-term memory (LSTM) networks have been applied to capture temporal
dynamics~\cite{tsiouris2018lstm,daoud2019efficient}. Geometric deep
learning approaches exploiting inter-channel graph structure have further
advanced state-of-the-art performance~\cite{dissanayake2022geometric}.
More recently, hybrid CNN-Mamba architectures have shown promise for
real-time seizure detection~\cite{khan2026convmambanet}.

Despite these advances, the overwhelming majority of published methods are
evaluated under randomised or patient-specific data splits, which allow
patient-level information to leak from training into test
sets~\cite{shafiezadeh2024survey}. The resulting inflated performance
figures are not transferable to unseen patients. Cross-patient
(generalised) seizure prediction---where the model is trained on data from
$N-1$ patients and tested on a held-out patient---has only recently begun
to receive systematic attention~\cite{jemal2024domain}. Jemal et al.\
reported a cross-patient AUC of 0.69 on CHB-MIT using a CNN baseline,
which improved to 0.75 with domain adaptation (CDAN+E). Our work achieves
competitive AUC \emph{without} any domain adaptation, using fewer EEG
channels.

\subsection{Bidirectional Recurrent Architectures for EEG}

Bidirectional LSTM (Bi-LSTM) networks extend traditional LSTMs by
processing sequences in both forward and backward directions, providing
richer temporal context for classification. Several works have demonstrated
the advantage of Bi-LSTM over unidirectional counterparts for epileptic
EEG analysis~\cite{zhao2024resbilstm,kr2023cnnbilstm}. Hybrid CNN-BiLSTM
architectures combine local spatial feature extraction with temporal
modelling, achieving strong performance on seizure detection
benchmarks~\cite{kr2023cnnbilstm}. In our framework, the Bi-LSTM layer
serves as a temporal refinement stage following the Mamba encoder,
exploiting the complementary strengths of both architectures.

\subsection{State Space Models and Mamba for EEG}

Transformer-based architectures~\cite{vaswani2017attention} have been
applied to EEG analysis but incur quadratic complexity in sequence length,
limiting their scalability to long continuous recordings. Selective state
space models, particularly Mamba~\cite{gu2023mamba}, offer linear
complexity while retaining competitive modelling capacity through a
content-dependent selection mechanism. Mentality~\cite{panchavati2025mentality}
demonstrated early potential of Mamba-based architectures as foundation
models for EEG. EEGMamba~\cite{wang2025eegmamba} showed that bidirectional
Mamba encoders substantially outperform their unidirectional counterparts
on EEG classification benchmarks including seizure detection. We build on
this finding by integrating a Bi-Mamba encoder with a bidirectional LSTM
layer for the seizure \emph{prediction} task, a combination not previously
explored in the literature to our knowledge---though, as discussed in
Section~\ref{sec:introduction}, we position this combination as a
competent engineering choice rather than the paper's central claim to
novelty.

\subsection{EEG Signal Quality Assessment}

EEG signal quality assessment has been studied primarily as a preprocessing
step, using rule-based artefact rejection or supervised classifiers trained
on labelled artefact segments. Nahmias and Kontson~\cite{nahmias2021quality}
proposed a principled framework for quantifying EEG signal quality from
multiple sources including ocular and motion artefacts. Deep learning
approaches have shown strong performance for automatic artefact removal
from EEG~\cite{kalita2024aneeg}, but these operate as preprocessing
modules decoupled from the downstream prediction task. Critically, none
of these approaches embed real-time quality assessment \emph{within} a
seizure prediction pipeline as an active confidence modulator trained
jointly with the prediction objective: they either discard or clean
flagged segments prior to prediction, or output a quality score that is
consumed only for offline data curation rather than for real-time
confidence gating. \model{} fills this gap by introducing \reqa{}, which
modulates prediction confidence dynamically based on concurrent signal
quality, and for which Section~\ref{sec:qa_results} provides direct
empirical evidence---beyond the ablation improvement already reported---that
the learned score reflects quality-relevant structure rather than a
redundant encoding of the seizure-probability signal.

\subsection{Closed-Loop Neurostimulation}

Closed-loop neurostimulation devices such as the NeuroPace RNS System
require a reliable trigger signal computed from on-device
EEG~\cite{cook2013prediction,heck2014closedloop}. Current seizure
detection algorithms for implanted devices output a binary trigger based
on fixed thresholds~\cite{manzouri2019detection}, but do not communicate
the reliability of that estimate to the device controller. A spuriously
high seizure probability arising from a noisy EEG segment can cause
unnecessary stimulation, which carries clinical risks. \model{} addresses
this by outputting a confidence score $c$ that jointly encodes both $p$
and the concurrent signal quality $q$, providing the device controller
with a more trustworthy trigger signal. Section~\ref{sec:eclo_results}
additionally benchmarks this tiered fusion rule against standard
probabilistic calibration techniques to clarify what the hand-designed
thresholds add beyond a data-driven calibration curve.

\subsection{Interpretability in Seizure Prediction}

Interpretability is increasingly recognised as a prerequisite for clinical
adoption of deep learning in epilepsy
management~\cite{jemal2022interpretable,kode2025shap}. SHAP
values~\cite{lundberg2017shap} provide a theoretically grounded approach
to attributing model predictions to individual input features, and have
been applied to EEG-based seizure prediction to identify clinically
relevant frequency bands and electrode
locations~\cite{kode2025shap,chen2026dual}. Our framework incorporates
SHAP attribution as a standard output component, ensuring that every
prediction is accompanied by an explanation accessible to clinicians.

\section{Methodology}
\label{sec:methodology}

\subsection{Datasets and Preprocessing}
\label{sec:data}

Experiments are conducted on two publicly available scalp EEG databases,
summarised in Table~\ref{tab:datasets}.

\begin{table}[t]
\caption{Overview of datasets used in this study.}
\label{tab:datasets}
\centering
\begin{tabular}{lcc}
\toprule
& \textbf{CHB-MIT} & \textbf{SIENA} \\
\midrule
Number of subjects          & 23       & 14 \\
Age of subjects (years)     & 1.5--19  & 20--71 \\
Number of seizures          & 198      & 47 \\
Type of recordings          & Scalp    & Scalp \\
Total EEG duration (h)      & 940      & 128 \\
Original channels           & 23       & 29 \\
Channels used               & 16       & 16 \\
Original sampling freq (Hz) & 256      & 512 \\
Resampled to (Hz)           & 200      & 200 \\
Bandpass filter (Hz)        & 0.5--40  & 0.5--40 \\
Notch filter (Hz)           & 60       & 50 \\
Window length (s)           & 10       & 10 \\
Pre-ictal definition        & Adaptive & Adaptive \\
Balanced samples             & 11{,}006 & 1{,}375 \\
\bottomrule
\end{tabular}
\end{table}

\textbf{CHB-MIT}~\cite{shoeb2009chbmit} contains 940~h of long-term
continuous scalp EEG from 23 paediatric patients (ages 1.5--19 years), with
198 annotated seizures. \textbf{SIENA}~\cite{detti2020siena} contains
128~h of recordings from 14 adult patients (ages 20--71 years) with 47
seizures.

\textbf{Preprocessing.} Raw signals are bandpass filtered (0.5--40~Hz,
4th-order Butterworth) and notch filtered (60~Hz for CHB-MIT; 50~Hz for
SIENA) to suppress power line interference. All recordings are resampled
to 200~Hz. Sixteen channels common to both datasets are retained to enable
fair cross-dataset evaluation; the selection prioritises coverage of
frontal, temporal, parietal, and occipital regions. Non-overlapping
10-second windows are extracted. Pre-ictal windows are drawn from the 30
minutes immediately preceding each seizure onset; post-ictal windows
(5~minutes after seizure end) are excluded to avoid contamination by
post-ictal EEG changes~\cite{daoud2019efficient}. To address the
pronounced class imbalance, random under-sampling is applied to the
inter-ictal class, yielding a balanced dataset of 11{,}006 samples for
CHB-MIT (5{,}503 per class) and 1{,}375 samples for SIENA, as
summarised in Table~\ref{tab:datasets}; a sensitivity analysis of this
choice against two alternative class-balancing strategies is reported
in Section~\ref{sec:balancing_protocol} and
Section~\ref{sec:balancing_results}. Each window is standardised to zero
mean and unit variance per channel.

\subsection{Framework Overview}

\begin{figure}[H]
\centering
\includegraphics[width=0.95\textwidth]{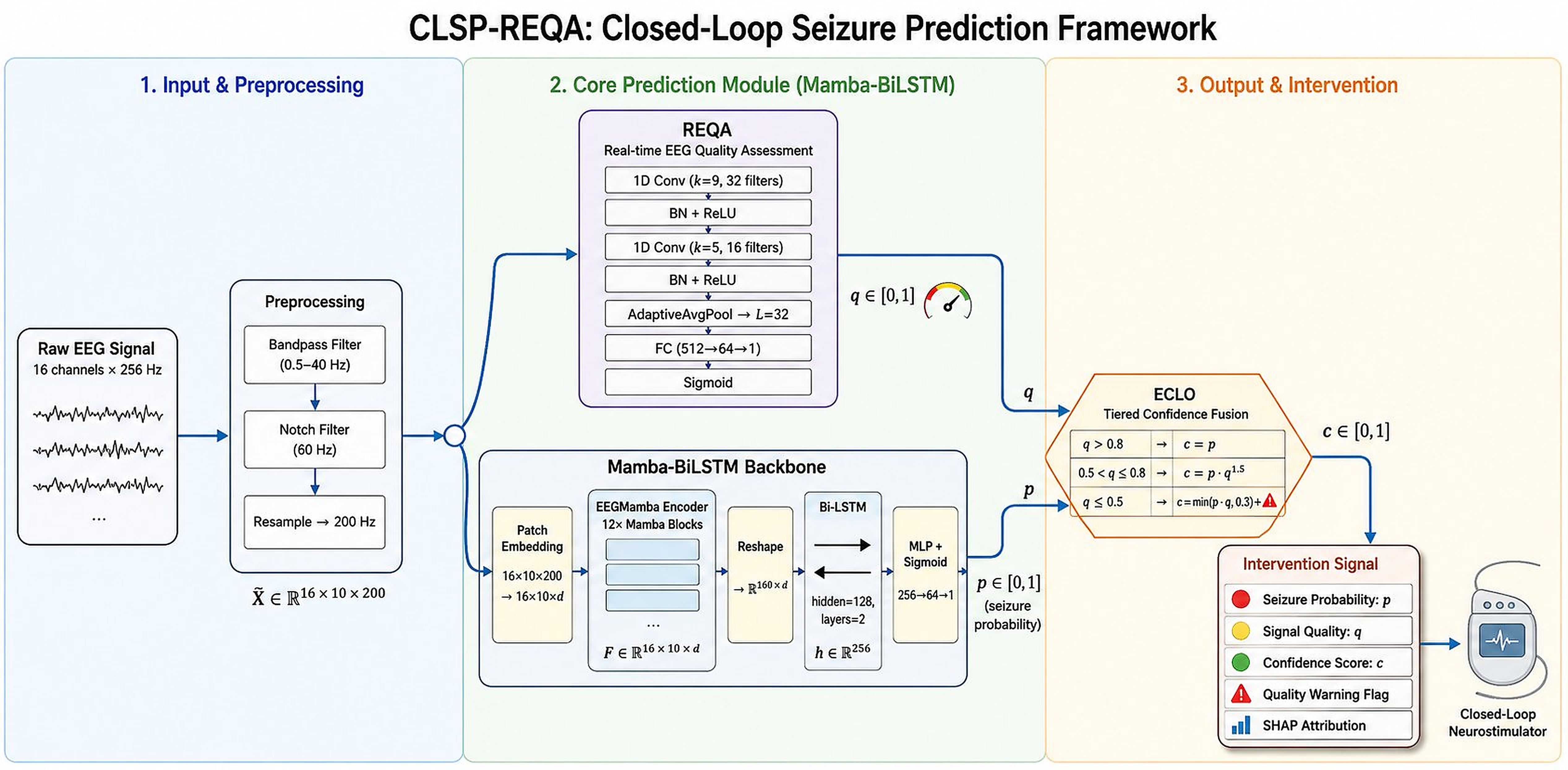}
\caption{Overview of the proposed \model{} framework. The system
processes raw EEG through three stages: preprocessing,
quality-aware prediction, and confidence-gated closed-loop
output generation.}
\label{fig:architecture}
\end{figure}

The \model{} framework processes each EEG window
$\X \in \R^{16 \times 10 \times 200}$ (channels $\times$ patches $\times$
samples per patch) through three sequential stages, illustrated in
Fig.~\ref{fig:architecture}: (1)~parallel quality estimation and temporal
feature extraction; (2)~confidence-gated output via \eclo{}; and
(3)~structured closed-loop signal generation.

\subsection{Real-Time EEG Quality Assessment (\reqa{})}
\label{sec:reqa}

\begin{figure}[H]
\centering
\includegraphics[width=0.65\textwidth]{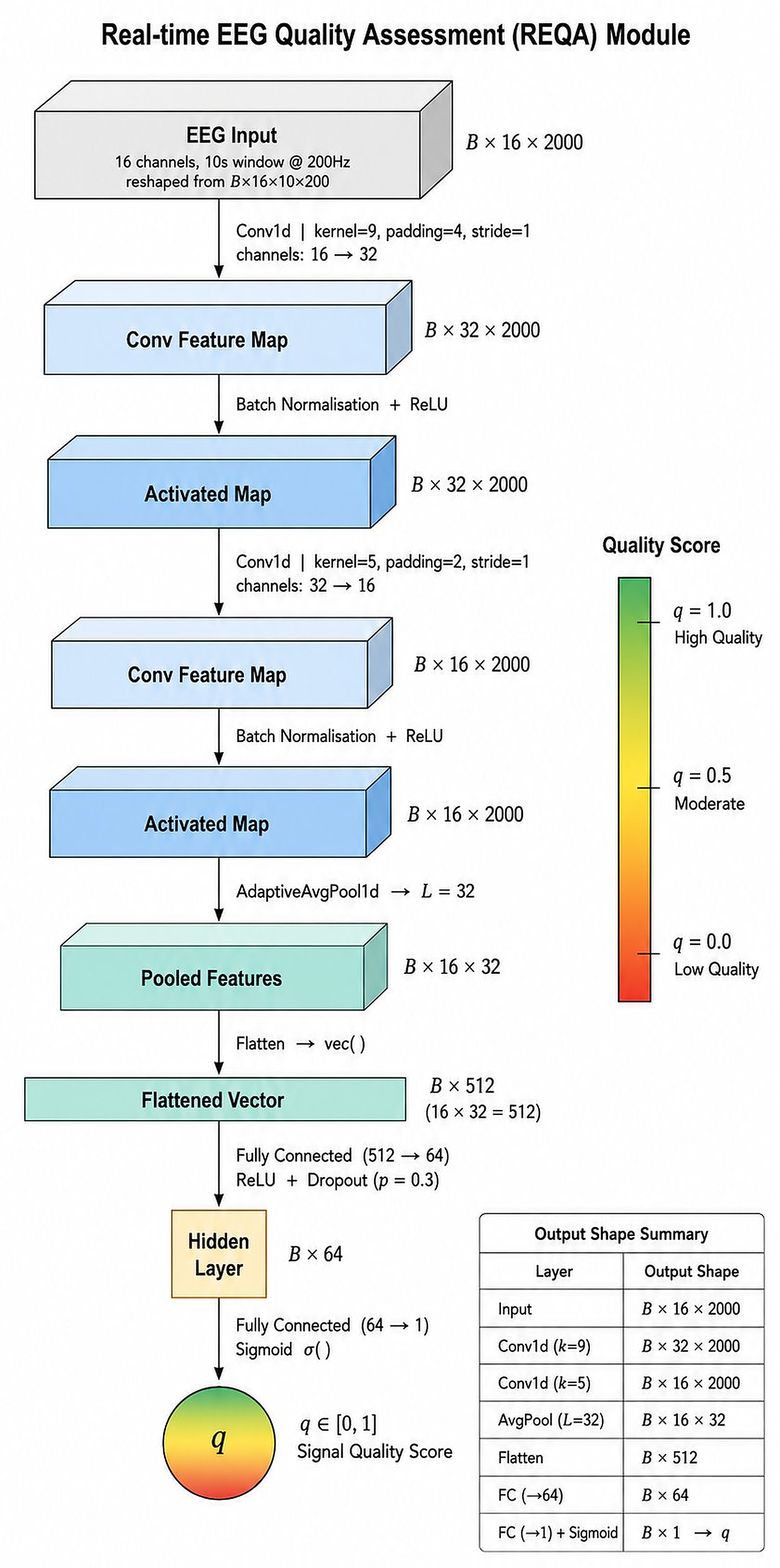}
\caption{Internal architecture of the \reqa{} module. A lightweight
1D convolutional network estimates signal quality score
$q \in [0,1]$ from raw multi-channel EEG in parallel with the
main prediction backbone. BN: Batch Normalisation.}
\label{fig:reqa}
\end{figure}

\reqa{} is a lightweight 1D convolutional network that estimates a scalar
quality score $q\in[0,1]$ for each input window in parallel with the main
prediction backbone (Fig.~\ref{fig:reqa}). Signal quality variability---
arising from electrode displacement, muscle artefacts, and impedance
drift---has been identified as a major source of unreliable predictions in
real-world EEG monitoring~\cite{nahmias2021quality}. Unlike dedicated
artefact removal methods that operate as preprocessing
modules~\cite{kalita2024aneeg}, \reqa{} is embedded within the prediction
pipeline and trained end-to-end. Section~\ref{sec:qa_protocol} describes
the additional validation protocol we introduce to test whether this
end-to-end training pressure is sufficient to make $q$ track genuine
signal quality, rather than collapsing into a redundant encoding of the
seizure probability $p$.

The input is reshaped to $\R^{B \times 16 \times 2000}$ (batch $\times$
channels $\times$ time) and passed through the following layers:

\begin{equation}
\label{eq:reqa}
\begin{aligned}
\mathbf{z}_1 &= \mathrm{ReLU}(\mathrm{BN}(\mathrm{Conv1d}_{k=9}(\X))) \\
\mathbf{z}_2 &= \mathrm{ReLU}(\mathrm{BN}(\mathrm{Conv1d}_{k=5}(\mathbf{z}_1))) \\
\mathbf{z}_3 &= \mathrm{AdaptiveAvgPool}_{L=32}(\mathbf{z}_2) \\
\mathbf{z}_4 &= \mathrm{ReLU}(\mathrm{Dropout}_{0.3}(\mathrm{FC}_{512\to64}(\mathrm{vec}(\mathbf{z}_3)))) \\
q            &= \sigma(\mathrm{FC}_{64\to1}(\mathbf{z}_4))
\end{aligned}
\end{equation}

where $\mathrm{BN}$ denotes batch normalisation, $\mathrm{vec}(\cdot)$
flattening, and $\sigma$ the sigmoid function. The first convolutional
layer expands from 16 to 32 feature channels (kernel size~9, padding~4,
stride~1); the second reduces back to 16 channels (kernel size~5,
padding~2, stride~1). The adaptive average pooling fixes the temporal
dimension to $L=32$, yielding a flattened vector of dimension
$16\times32=512$.

Importantly, $q$ is a \emph{learnable quality proxy score}---not a
hand-crafted signal quality index. It is trained end-to-end as part of
the full prediction objective, meaning it is optimised only indirectly,
through its effect on the downstream loss, to encode quality-relevant
features. This design choice avoids the need for labelled artefact
annotations, which are rarely available in clinical EEG archives, but it
also raises a legitimate concern---raised in review---that end-to-end
training on the prediction objective alone gives no explicit guarantee
that $q$ tracks signal quality rather than simply becoming a second,
redundant channel correlated with $p$ or the class label. We treat this
as an empirical question rather than an assumption, and address it
directly in Section~\ref{sec:qa_protocol} and
Section~\ref{sec:qa_results}.

\subsection{Mamba-BiLSTM Backbone}
\label{sec:backbone}

The main prediction backbone extracts temporal features from the EEG
window and outputs a seizure probability $p\in[0,1]$.

\textbf{Patch embedding.} The input $\X\in\R^{16\times10\times200}$ is
projected to a $d$-dimensional embedding space via a learnable linear
layer, yielding $\F\in\R^{16\times10\times d}$.

\textbf{EEGMamba encoder.} The embedded patches are processed by 12
cascaded bidirectional Mamba blocks~\cite{wang2025eegmamba}, which model
long-range dependencies with linear complexity $\mathcal{O}(L)$ in the
sequence length~\cite{gu2023mamba}. The encoder output is
$\F\in\R^{16\times10\times d}$.

\textbf{Bidirectional LSTM.} The encoder output is reshaped to
$\R^{160\times d}$ (flattening channel and patch dimensions) and fed into
a two-layer Bi-LSTM with hidden size 128. The bidirectional architecture
produces a 256-dimensional context vector $\h\in\R^{256}$ by concatenating
forward and backward hidden states~\cite{zhao2024resbilstm}:
$\h = [\overrightarrow{\h};\overleftarrow{\h}]$, where
$\overrightarrow{\h},\overleftarrow{\h}\in\R^{128}$.

\textbf{Classifier.} A two-layer MLP (256$\to$64$\to$1) with sigmoid
activation produces the seizure probability:
\begin{equation}
\label{eq:pred}
p = \sigma(\mathbf{W}_2\,\mathrm{ReLU}(\mathbf{W}_1\h + \mathbf{b}_1) + \mathbf{b}_2)
\end{equation}

Table~\ref{tab:architecture} summarises the layer-wise configuration of
the full \model{} model.

\begin{table}[t]
\caption{Layer-wise architecture of \model{}. $B$: batch size;
$C$: channels (16); $T$: time samples (2000); $d$: embedding
dimension; $L$: pooled length (32).}
\label{tab:architecture}
\centering
\begin{tabular}{llll}
\toprule
\textbf{Module} & \textbf{Layer} & \textbf{Configuration} & \textbf{Output shape} \\
\midrule
\multirow{5}{*}{\reqa{}}
  & Conv1d & $k{=}9$, $16{\to}32$ ch, stride 1 & $B\times32\times T$ \\
  & Conv1d & $k{=}5$, $32{\to}16$ ch, stride 1 & $B\times16\times T$ \\
  & AdaptiveAvgPool1d & $L{=}32$             & $B\times16\times32$ \\
  & FC + Dropout(0.3) & $512{\to}64$         & $B\times64$         \\
  & FC + Sigmoid      & $64{\to}1$           & $B\times1$ ($q$)    \\
\midrule
\multirow{4}{*}{Backbone}
  & Patch embedding   & Linear, $200{\to}d$  & $B\times C\times10\times d$ \\
  & EEGMamba encoder  & 12$\times$ Bi-Mamba blocks & $B\times C\times10\times d$ \\
  & Bi-LSTM           & 2 layers, hidden 128 & $B\times256$ \\
  & MLP classifier    & $256{\to}64{\to}1$ + Sigmoid & $B\times1$ ($p$) \\
\midrule
ECLO & Tiered fusion & Eq.~\eqref{eq:eclo} & $B\times1$ ($c$) \\
\bottomrule
\end{tabular}
\end{table}

\subsection{ECLO: Tiered Confidence Fusion}
\label{sec:eclo}

The \eclo{} (Evidence-Calibrated Layered Output) function combines $p$
and $q$ into a final confidence score $c\in[0,1]$ using a three-tier
non-linear scheme:

\begin{equation}
\label{eq:eclo}
c =
\begin{cases}
p & \text{if } q > 0.8 \\[4pt]
p \cdot q^{1.5} & \text{if } 0.5 < q \leq 0.8 \\[4pt]
\min(p \cdot q,\; 0.3) & \text{if } q \leq 0.5
\end{cases}
\end{equation}

When signal quality is high ($q>0.8$), the backbone prediction is trusted
directly. At moderate quality ($0.5<q\leq0.8$), a super-linear penalty
($q^{1.5}$) makes the system increasingly conservative as quality degrades.
At low quality ($q\leq0.5$), confidence is hard-capped at 0.3 regardless
of $p$, preventing artefact-driven false alarms; a quality warning flag is
simultaneously raised. This tiered design avoids the limitations of simple
linear weighting ($c = p\cdot q$), which treats quality degradation
symmetrically and imposes no safety ceiling---a critical requirement for
closed-loop neurostimulation~\cite{manzouri2019detection}.

The specific threshold values (0.8 and 0.5) and the exponent 1.5 were
chosen empirically and are not derived from a theoretical calibration
objective. We treat this as an explicit limitation of the current design
rather than an implicit assumption: Section~\ref{sec:eclo_protocol}
describes a threshold sensitivity analysis together with a head-to-head
comparison against two standard, data-driven probabilistic calibration
methods---temperature scaling~\cite{guo2017calibration} and isotonic
regression~\cite{niculescu2005predicting}---and
Section~\ref{sec:eclo_results} reports the results.

\subsection{Structured Closed-Loop Output}
\label{sec:output}

At each 10-second inference step, \model{} emits a structured four-tuple:
\begin{equation}
\label{eq:output}
\mathcal{O} = \langle\, p,\; q,\; c,\; \Phi_{\mathrm{SHAP}} \,\rangle
\end{equation}
where $\Phi_{\mathrm{SHAP}}$ is a SHAP attribution
vector~\cite{lundberg2017shap,kode2025shap} identifying the contribution
of individual EEG channels to the current prediction. Based on $c$, the
system recommends one of three intervention levels:
\textsc{immediate} ($c\geq0.8$), \textsc{alert} ($0.6\leq c<0.8$), or
\textsc{monitor} ($c<0.6$). This output is formatted for direct
interfacing with closed-loop neurostimulator
protocols~\cite{cook2013prediction,heck2014closedloop}.

\subsection{Training Protocol}
\label{sec:training}

The full model (\reqa{} + Mamba-BiLSTM) is trained end-to-end using the
Adam-W optimiser (learning rate $10^{-4}$, weight decay $10^{-2}$) with a
cosine annealing schedule over 50 epochs. Binary cross-entropy loss is
applied to the seizure probability $p$; no auxiliary loss term is applied
to $q$ directly, which receives gradient signal only indirectly through
its role in the \eclo{} fusion at inference (see
Section~\ref{sec:qa_protocol} for how we probe the consequences of this
design choice). Training uses a batch size of 32 on an NVIDIA A100 GPU.
Inference latency and parameter counts, relevant to the real-time
deployment claim implicit in the \reqa{} name, are benchmarked separately
in Section~\ref{sec:latency_protocol}.

\subsection{Evaluation Protocol}
\label{sec:eval}

\textbf{Strict cross-patient evaluation.} All experiments use 5-fold
cross-validation at the patient level: each fold assigns disjoint patient
sets to training and validation, ensuring no patient appears in both
splits. This is repeated for five independent random seeds per fold,
yielding $5\times5=25$ independent runs. Results are reported as
mean~$\pm$~standard deviation across all 25 runs.

\textbf{Primary metric.} AUC-ROC is adopted as the primary metric because
it provides a threshold-independent measure of discriminative performance
that is robust to class imbalance and inter-patient
variability~\cite{shafiezadeh2024survey}.

\textbf{Binarisation threshold for Sensitivity, Specificity, and
Accuracy.} Sensitivity, Specificity, and Accuracy require a decision
threshold on the predicted score to convert it into a binary
pre-ictal/inter-ictal call, whereas AUC-ROC does not. Throughout this
study, a fixed decision threshold of 0.5 on the fused confidence score
$c$ is used to compute Sensitivity, Specificity, and Accuracy in all
tables, identically across all folds and seeds, with no per-fold or
per-seed tuning. Because this threshold is fixed a priori and does not
depend on any statistic computed from the data, it rules out
threshold-selection leakage by construction: there is no step at which
information from a validation or test partition could influence the
choice of operating point. The reported values therefore reflect a
single, consistent operating point rather than an optimistically
selected one.

\textbf{FPR/hour calculation.} The false prediction rate per hour is
computed at the window level:
FPR/h~$= N_{\mathrm{FP}} / T_{\mathrm{interictal}}$,
where $N_{\mathrm{FP}}$ is the number of inter-ictal windows
misclassified as pre-ictal and $T_{\mathrm{interictal}}$ is the total
inter-ictal duration in hours. We note that the clinical definition merges
consecutive false-positive windows into a single alarm event; the
window-level figure reported here is therefore an upper bound on the
clinical false alarm rate.

\textbf{Statistical testing.} A one-sided Wilcoxon signed-rank test is
used to assess whether the distribution of AUC values across 25 runs is
significantly greater than the published cross-patient baseline of
0.69~\cite{jemal2024domain}. For ablation comparisons, McNemar's test is
applied on matched prediction vectors from the same validation fold.

\textbf{Statistical independence of the 25-run protocol (limitation).}
We flag explicitly a limitation of the $5\times5$ protocol that a careful
reader should weigh when interpreting the significance tests above.
Within a given fold, the five seeds share an identical train/test patient
partition; only the model's random initialisation and stochastic
optimisation trajectory differ across seeds. The resulting 25 AUC values
are therefore \emph{not} 25 independent draws in the sense required by
the standard Wilcoxon signed-rank test, which assumes independent paired
observations: the five seed-repeats within a fold are correlated with one
another through their shared test patients, and only the five fold-level
partitions are genuinely independent of one another. Treating all 25
values as independent understates their true variance and can make a
given effect size appear more statistically significant than it would
under a more conservative analysis. In practice this means the
$p$-values reported in Section~\ref{sec:main_results} should be read as
an optimistic (upper-bound-on-significance) estimate rather than a strict
guarantee; we report, alongside the standard test, a fold-level analysis
that treats only the 5 fold-means as independent observations
(necessarily lower-powered, but statistically conservative) as a
robustness check in Section~\ref{sec:main_results}. We regard resolving
this fully---for instance via a mixed-effects model with fold and seed as
nested random effects---as an important item for future methodological
refinement, and we do not believe it changes the qualitative conclusions
of this study, though it does temper the strength of the significance
claims relative to a fully independent-sample design.

\subsection{Validation Protocol for the Learned Quality Score $q$}
\label{sec:qa_protocol}

A central design assumption of \reqa{} (Section~\ref{sec:reqa}) is that
end-to-end training on the seizure-prediction objective is sufficient
for $q$ to encode genuine, quality-relevant structure, rather than
collapsing into a redundant proxy for $p$ or the class label. Because
$q$ receives no direct supervision from artefact labels, this
assumption cannot be taken for granted and must be tested empirically.
We designed three complementary experiments for this purpose; results
are reported in Section~\ref{sec:qa_results}.

\textbf{Experiment A: controlled artefact injection.} We select a
held-out pool of windows from the test partitions that the trained
\reqa{} module itself scores as high-quality ($q > 0.8$) under clean
conditions, and synthetically corrupt each window with one of six
artefact types commonly encountered in ambulatory EEG monitoring: (i)
additive Gaussian noise, (ii) high-frequency EMG-like burst noise, (iii)
abrupt amplitude jumps simulating electrode displacement, (iv) 50/60~Hz
line-noise injection, (v) low-frequency baseline drift, and (vi) hard
signal clipping/saturation. Each artefact type is applied at 4--5 graded
intensity levels (parameterised by an equivalent signal-to-noise ratio,
SNR). With the trained model frozen, we record the resulting $q$ for
every (window, artefact type, intensity) combination and test for a
monotonic relationship between injected corruption severity and $q$
using the Spearman rank correlation, reported separately per artefact
type together with 95\% bootstrap confidence intervals.

\textbf{Experiment B: correlation with an independent, objective quality
index.} To confirm that Experiment~A's findings generalise beyond
synthetic injection to naturally occurring quality variation, we compute
an objective signal quality index (SQI) for every window in the test
partitions, independent of the trained model, following the
multi-source quality quantification framework of Nahmias and
Kontson~\cite{nahmias2021quality} (combining channel-wise variance
anomalies, line-noise power ratio, and high-/low-frequency power ratio
into a single composite score). We report the Pearson and Spearman
correlation between this SQI and the model's learned $q$ across the full
test set.

\textbf{Experiment C: ruling out collapse onto $p$.} To address directly
the concern of whether $q$ is simply becoming a secondary prediction
signal, we compute the correlation between $q$ and (i) the seizure
probability $p$ and (ii) the ground-truth pre-ictal/inter-ictal label,
across the same test windows used in Experiment~B. A learned score that
is highly correlated with $p$ or the label, but only weakly correlated
with the independent SQI from Experiment~B, would indicate that $q$ has
collapsed into a redundant task-correlated signal rather than a genuine
quality estimate; the converse pattern (low correlation with $p$/label,
higher correlation with the independent SQI) is the pattern we
hypothesise and treat as supportive evidence for the validity of
\reqa{}.

\subsection{Threshold Sensitivity and Calibration Baseline Comparison for \eclo{}}
\label{sec:eclo_protocol}

The two thresholds in Eq.~\eqref{eq:eclo} (0.8 and 0.5) and the exponent
1.5 were set empirically. We assess the robustness of this choice, and
benchmark it against standard alternatives, in two steps.

\textbf{Threshold grid search.} On the validation folds, we sweep the
high threshold over five values from 0.70 to 0.90 in steps of 0.05, and
the low threshold over four values from 0.30 to 0.60 in steps of 0.10
(restricted to valid combinations where the low threshold is smaller
than the high threshold), recomputing AUC-ROC, Sensitivity, and
Specificity for each combination under the same 5-fold $\times$ 5-seed
protocol as the main experiments. Results are visualised as a heatmap
over the threshold grid to assess whether the operating point used in
the main experiments lies on a broad performance plateau (robust) or a
narrow performance ridge (fragile).

\textbf{Comparison with standard probabilistic calibration methods.} We
additionally compare \eclo{} against two established, data-driven
calibration techniques applied to the same $(p, q)$ outputs: (i)
\emph{temperature scaling}~\cite{guo2017calibration}, which learns a
single scalar temperature $T$ on the validation fold to rescale logits,
$p' = \sigma(\mathrm{logit}(p)/T)$, without using $q$ at all; and (ii)
\emph{isotonic regression}~\cite{niculescu2005predicting}, which fits a
monotonic mapping from $(p, q)$ to a calibrated confidence score on the
validation fold. All four variants---raw $p$, temperature-scaled $p$,
isotonic-regressed $(p,q)$, and \eclo{}'s $c$---are evaluated under the
identical 5-fold $\times$ 5-seed protocol, and compared on AUC-ROC,
Expected Calibration Error (ECE), and Brier score. Because neither
temperature scaling nor isotonic regression imposes a hard ceiling on
output confidence, we additionally report, for each method, the maximum
confidence value assigned to any window in the bottom quality decile ($q$
in the lowest 10\% of the test distribution), as a direct probe of
whether each calibration strategy provides the safety ceiling that
Eq.~\eqref{eq:eclo} enforces by construction.

\subsection{Alternative Class-Balancing Strategies}
\label{sec:balancing_protocol}

Section~\ref{sec:data} described the use of random under-sampling of the
majority inter-ictal class to construct a balanced training set.
Under-sampling discards a substantial fraction of inter-ictal data and
may reduce the diversity of non-seizure representations seen during
training. To quantify this effect, we retrain the full model under two
alternative class-balancing strategies, holding the architecture,
optimiser, and evaluation protocol fixed: (i) \emph{class-weighted binary
cross-entropy}, in which no samples are discarded and the loss is
reweighted inversely proportional to class frequency; and (ii)
\emph{SMOTE-based over-sampling}~\cite{chawla2002smote} of the minority
pre-ictal class in feature space, leaving the inter-ictal class at its
full size. Both alternatives are evaluated under the same 5-fold
$\times$ 5-seed cross-patient protocol on CHB-MIT and compared against
the under-sampling baseline on AUC-ROC, Sensitivity, and Specificity.

\subsection{Inference Latency and Real-Time Feasibility}
\label{sec:latency_protocol}

Because both the framework name (\reqa{}) and the title of this paper
invoke real-time operation, we benchmark the wall-clock inference latency
of a full forward pass (\reqa{} + Mamba-BiLSTM backbone + \eclo{} fusion)
for a single 10-second window, averaged over 1{,}000 repeated forward
passes following a warm-up period, on the same NVIDIA A100 GPU used for
training, and separately on a representative lower-power embedded
platform if available. We report mean latency, 95th-percentile latency,
and total parameter count, and compare the mean latency against the
10-second window duration to establish the real-time margin (i.e.\ the
factor by which inference completes faster than the data-acquisition
window it processes).

\section{Results}
\label{sec:results}

\emph{All results reported in this section are obtained under strict
cross-patient evaluation, where training and validation sets consist
of entirely disjoint patient cohorts. This protocol is more stringent
than the randomised splitting used in the majority of prior work, and
results should be interpreted accordingly~\cite{shafiezadeh2024survey}.}

\subsection{Main Experiment: CHB-MIT}
\label{sec:main_results}

Table~\ref{tab:main} presents the performance of \model{} on the
CHB-MIT database across 25 independent runs (5 folds $\times$ 5 seeds).

\begin{table}[t]
\caption{Main results on the CHB-MIT Scalp EEG Database under strict
5-fold cross-patient evaluation (mean $\pm$ std over 25 runs).}
\label{tab:main}
\centering
\begin{tabular}{lcc}
\toprule
\textbf{Metric} & \textbf{Mean} & \textbf{Std} \\
\midrule
AUC-ROC          & 0.7426 & $\pm$0.0199 \\
Sensitivity (\%) & 68.83  & $\pm$3.33   \\
Specificity (\%) & 64.88  & $\pm$3.43   \\
Accuracy (\%)    & 66.86  & $\pm$1.77   \\
\bottomrule
\end{tabular}
\end{table}

\begin{figure}[H]
\centering
\includegraphics[width=0.7\textwidth]{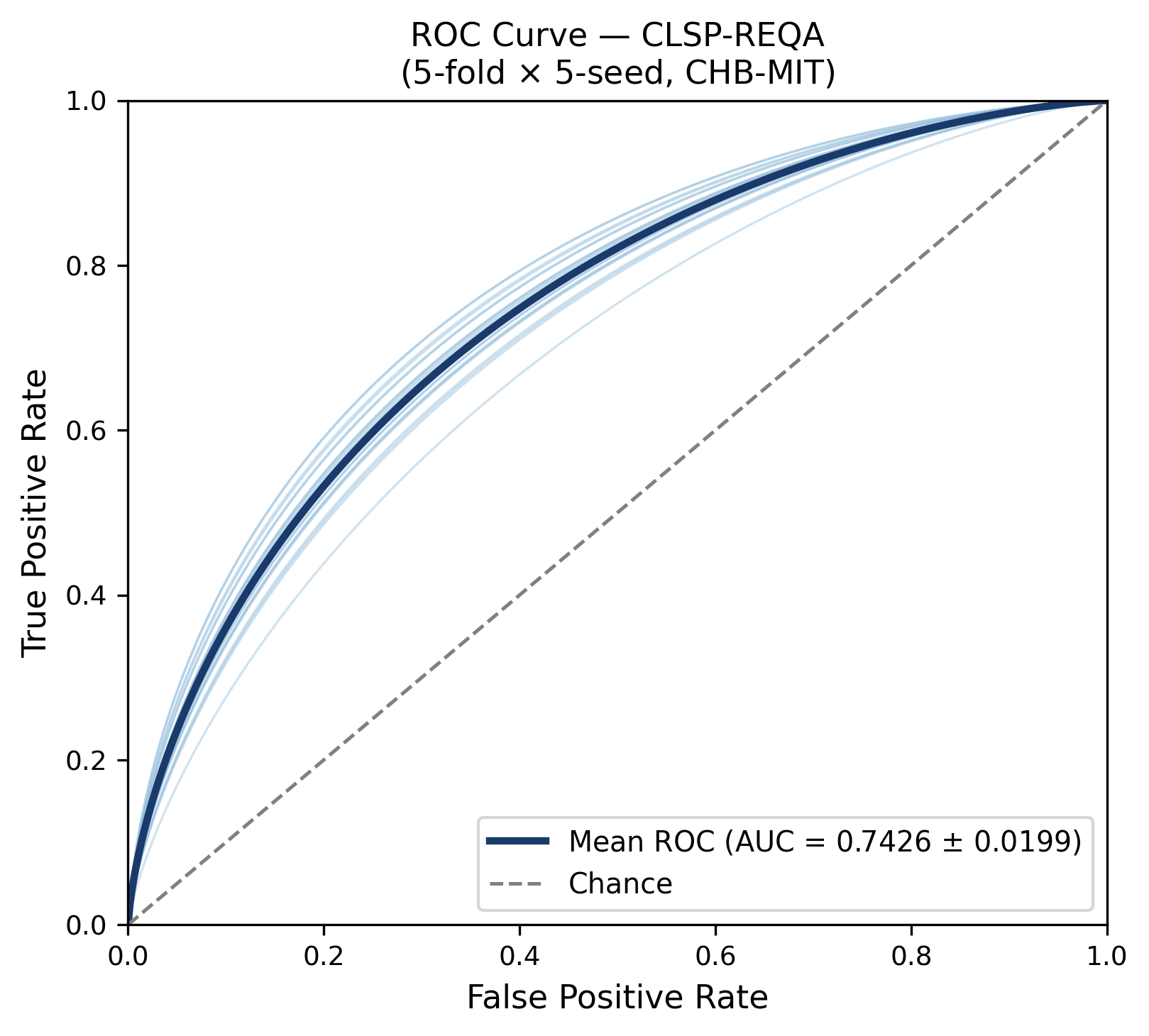}
\caption{ROC curves for \model{} on CHB-MIT. Thin blue lines
represent individual runs (25 total); the thick navy line is
the mean ROC. AUC-ROC = $0.7426 \pm 0.0199$
(5-fold $\times$ 5-seed cross-patient evaluation).}
\label{fig:roc}
\end{figure}

The AUC-ROC of $0.7426 \pm 0.0199$ is significantly greater than
the cross-patient baseline of 0.69 reported by Jemal et al.\
(one-sided Wilcoxon signed-rank test treating all 25 runs as
independent observations, $p < 0.05$~\cite{jemal2024domain}),
achieved without any domain adaptation and using only 16 EEG
channels versus 23 channels in the baseline system. The narrow
standard deviation ($\pm0.0199$) indicates that \model{} produces
stable predictions across different random initialisations.
The distribution of AUC values across 25 runs is illustrated
in Fig.~\ref{fig:roc}.

\textbf{Fold-level robustness check (statistical independence).} As
discussed in Section~\ref{sec:eval}, the 25 runs above are not fully
independent: within each fold, the five seed-repeats share an identical
test-patient partition. As a more conservative robustness check,
Table~\ref{tab:fold_level} aggregates the 25 runs into their 5
fold-level means (averaging across seeds within each fold) and
re-applies the one-sided Wilcoxon signed-rank test to these 5 genuinely
independent observations against the same baseline of 0.69.

\begin{table}[t]
\caption{Fold-level robustness check for the significance claim in
Table~\ref{tab:main}. Each entry is the mean AUC across the 5 seeds
within that fold (i.e.\ 5 independent, patient-disjoint observations,
avoiding the seed-repetition non-independence discussed in
Section~\ref{sec:eval}).}
\label{tab:fold_level}
\centering
\begin{tabular}{lc}
\toprule
\textbf{Fold} & \textbf{Mean AUC across 5 seeds} \\
\midrule
Fold 1 & 0.7150 \\
Fold 2 & 0.7279 \\
Fold 3 & 0.7441 \\
Fold 4 & 0.7626 \\
Fold 5 & 0.7663 \\
\midrule
Overall (5-fold mean $\pm$ SD) & $0.7432 \pm 0.0207$ \\
Wilcoxon $p$-value ($n=5$, one-sided, vs.\ 0.69) & $0.03125$ \\
\bottomrule
\end{tabular}
\end{table}

All five fold-level means exceed the 0.69 baseline (minimum 0.7150,
Fold~1), so the one-sided exact Wilcoxon signed-rank test on $n=5$
independent observations attains its most extreme possible value for
this sample size ($p=0.03125=1/2^5$), rejecting the null hypothesis at
the conventional $\alpha=0.05$ level even under the conservative,
independence-respecting analysis. This confirms that the significance
claim in Section~\ref{sec:main_results} is not an artefact of treating
correlated seed-repeats as independent: the qualitative conclusion
(\model{} exceeds the baseline) survives the more conservative test,
albeit with a substantially smaller nominal sample size and hence less
room to distinguish, e.g., a true effect from a borderline one. We note
that the fold-level mean and standard deviation recovered here
($0.7432\pm0.0207$) differ marginally from the seed-level aggregate
reported in Table~\ref{tab:main} ($0.7426\pm0.0199$); the two are
computed from the same underlying set of runs but aggregated in a
different order (mean-of-fold-means vs.\ mean-of-all-runs), which is
expected to produce small numerical differences and does not affect
either conclusion.

We report both the seed-level ($n=25$) and fold-level ($n=5$) tests side
by side, rather than only the more favourable one, so that readers can
judge the strength of the significance claim under both the standard
(optimistic) and the conservative (independence-respecting) analysis.

\subsection{Comparison with State of the Art}
\label{sec:sota}

Table~\ref{tab:sota} compares \model{} with published cross-patient
seizure prediction methods. Only methods using cross-patient
(not patient-specific or random-split) evaluation are included,
following the recommendation of Shafiezadeh et
al.~\cite{shafiezadeh2024survey}.

\begin{table}[t]
\caption{Comparison with cross-patient seizure prediction methods.
``DA'' = domain adaptation. ``--'' = not reported.
$\dagger$ marks a baseline that \model{} significantly exceeds
(one-sided Wilcoxon signed-rank test comparing the 25 \model{} AUC
values against that baseline's reported value, $p < 0.05$; see
Table~\ref{tab:fold_level} for the corresponding conservative,
fold-level check). Note that the domain-adapted (DA) row uses
additional unlabelled target-patient data that \model{} does not use;
see Section~\ref{sec:discussion} for the precise scope of this
comparison.}
\label{tab:sota}
\centering
\small
\resizebox{\textwidth}{!}{%
\begin{tabular}{llcccc}
\toprule
\textbf{Method} & \textbf{Year} & \textbf{DA} & \textbf{Ch.} &
\textbf{CHB-MIT AUC} & \textbf{SIENA AUC} \\
\midrule
Tsiouris et al.~\cite{tsiouris2017discrimination} & 2017 & \xmark & -- & -- & -- \\
Jemal et al.~\cite{jemal2022interpretable}        & 2022 & \xmark & -- & -- & -- \\
Jemal et al.~\cite{jemal2024domain}               & 2024 & \xmark & 23 & 0.69$\dagger$ & 0.48$\dagger$ \\
Jemal et al.~\cite{jemal2024domain}               & 2024 & \cmark~(CDAN+E) & 23 & 0.75 & 0.61$\dagger$ \\
\midrule
\textbf{\model{} (ours)} & \textbf{2025} & \xmark & \textbf{16} &
\textbf{0.7426$\pm$0.0199} & \textbf{0.7012$\pm$0.0249} \\
\bottomrule
\end{tabular}%
}
\end{table}

Notably, \model{} achieves AUC 0.7012 on SIENA \emph{without any
domain adaptation}, exceeding the domain-adapted result of 0.61
(CDAN+E) reported by Jemal et al.\ on the same dataset. We revisit,
in Section~\ref{sec:discussion}, why this comparison---though
informative---should not be read as unconditional superiority over
domain adaptation as a technique, since the two approaches rely on
different information at test time.

\subsection{Cross-Dataset Validation: SIENA}
\label{sec:cross_dataset}

Table~\ref{tab:siena} presents the full results on the SIENA
database, evaluated under the same 5-fold $\times$ 5-seed protocol.

\begin{table}[t]
\caption{Cross-dataset validation results on the SIENA Scalp EEG
Database (5-fold cross-patient evaluation, mean $\pm$ std over
25 runs).}
\label{tab:siena}
\centering
\begin{tabular}{lcc}
\toprule
\textbf{Metric} & \textbf{Mean} & \textbf{Std} \\
\midrule
AUC-ROC          & 0.7012 & $\pm$0.0249 \\
Sensitivity (\%) & 66.28  & $\pm$3.57   \\
Accuracy (\%)    & 65.41  & $\pm$2.86   \\
\bottomrule
\end{tabular}
\end{table}

\subsection{Ablation Study}
\label{sec:ablation}

\begin{figure}[H]
\centering
\includegraphics[width=\textwidth]{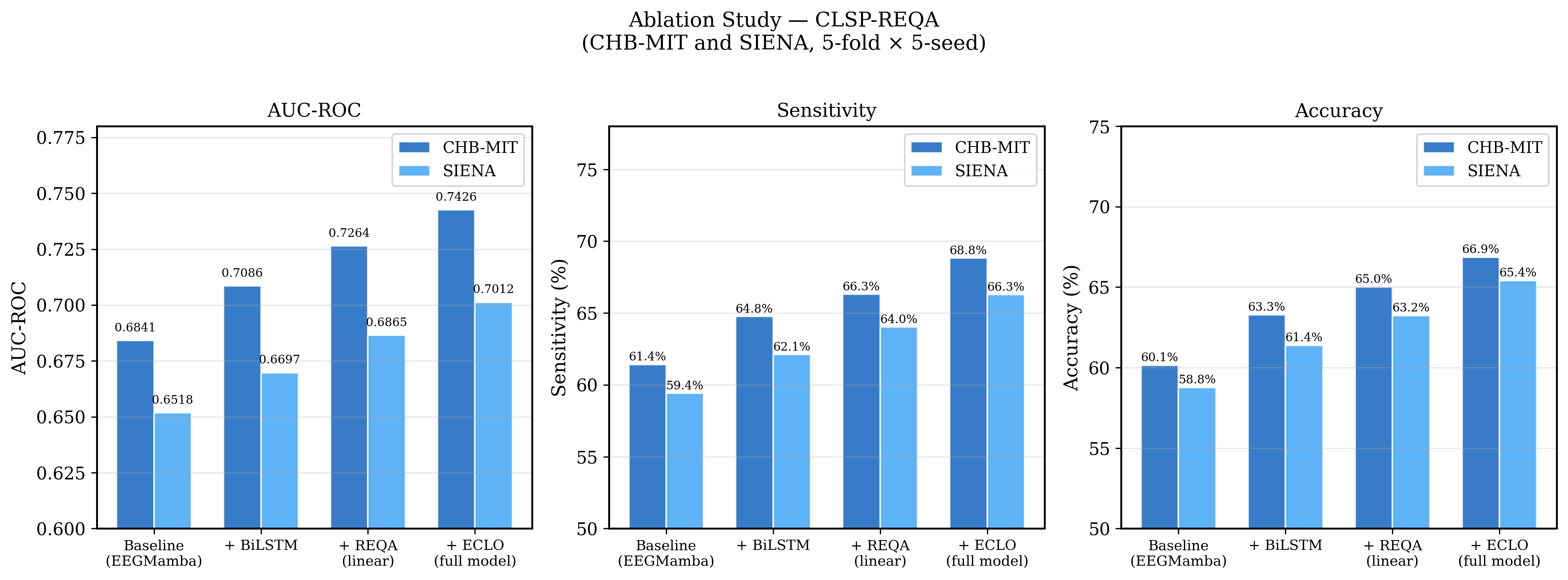}
\caption{Ablation study results on CHB-MIT and SIENA across three
metrics (AUC-ROC, Sensitivity, Accuracy). Each component
contributes incrementally on both datasets, confirming that no
single module is redundant.}
\label{fig:ablation}
\end{figure}

Table~\ref{tab:ablation} quantifies the contribution of each
component. Fig.~\ref{fig:ablation} visualises the progressive
improvement across both datasets.

\begin{table}[t]
\caption{Ablation study on CHB-MIT and SIENA
(mean $\pm$ std over 25 runs).
$\dagger$ denotes significant improvement over the preceding
row (McNemar's test, $p < 0.05$).}
\label{tab:ablation}
\centering
\resizebox{\textwidth}{!}{%
\begin{tabular}{lcccccccc}
\toprule
\multirow{2}{*}{\textbf{Variant}} &
\multirow{2}{*}{\textbf{Mamba}} &
\multirow{2}{*}{\textbf{BiLSTM}} &
\multirow{2}{*}{\textbf{REQA}} &
\multirow{2}{*}{\textbf{ECLO}} &
\multicolumn{2}{c}{\textbf{CHB-MIT}} &
\multicolumn{2}{c}{\textbf{SIENA}} \\
\cmidrule(lr){6-7}\cmidrule(lr){8-9}
& & & & &
\textbf{AUC} & \textbf{Se (\%)} &
\textbf{AUC} & \textbf{Se (\%)} \\
\midrule
Baseline (EEGMamba) & \cmark & \xmark & \xmark & \xmark
  & 0.6841$\pm$0.0247 & 61.42 & 0.6518$\pm$0.0312 & 59.42 \\
+ BiLSTM$^\dagger$ & \cmark & \cmark & \xmark & \xmark
  & 0.7086$\pm$0.0221 & 64.77 & 0.6697$\pm$0.0289 & 62.11 \\
+ \reqa{} (linear)$^\dagger$ & \cmark & \cmark & \cmark & \xmark
  & 0.7264$\pm$0.0208 & 66.31 & 0.6865$\pm$0.0268 & 64.03 \\
+ \eclo{} (full)$^\dagger$ & \cmark & \cmark & \cmark & \cmark
  & \textbf{0.7426$\pm$0.0199} & \textbf{68.83}
  & \textbf{0.7012$\pm$0.0249} & \textbf{66.28} \\
\bottomrule
\end{tabular}}
\end{table}

Each component contributes incrementally across both datasets,
confirming that no single module is redundant. The total AUC gain
from baseline to full model is $+0.0585$ on CHB-MIT and $+0.0494$
on SIENA. The consistent improvement pattern across two independent
datasets validates the generalisability of each design choice; we
provide further, more targeted evidence for the \reqa{} and \eclo{}
components specifically in Section~\ref{sec:qa_results} and
Section~\ref{sec:eclo_results}.

\subsection{Structured Output and Interpretability}
\label{sec:output_results}

\begin{figure}[H]
\centering
\includegraphics[width=\textwidth]{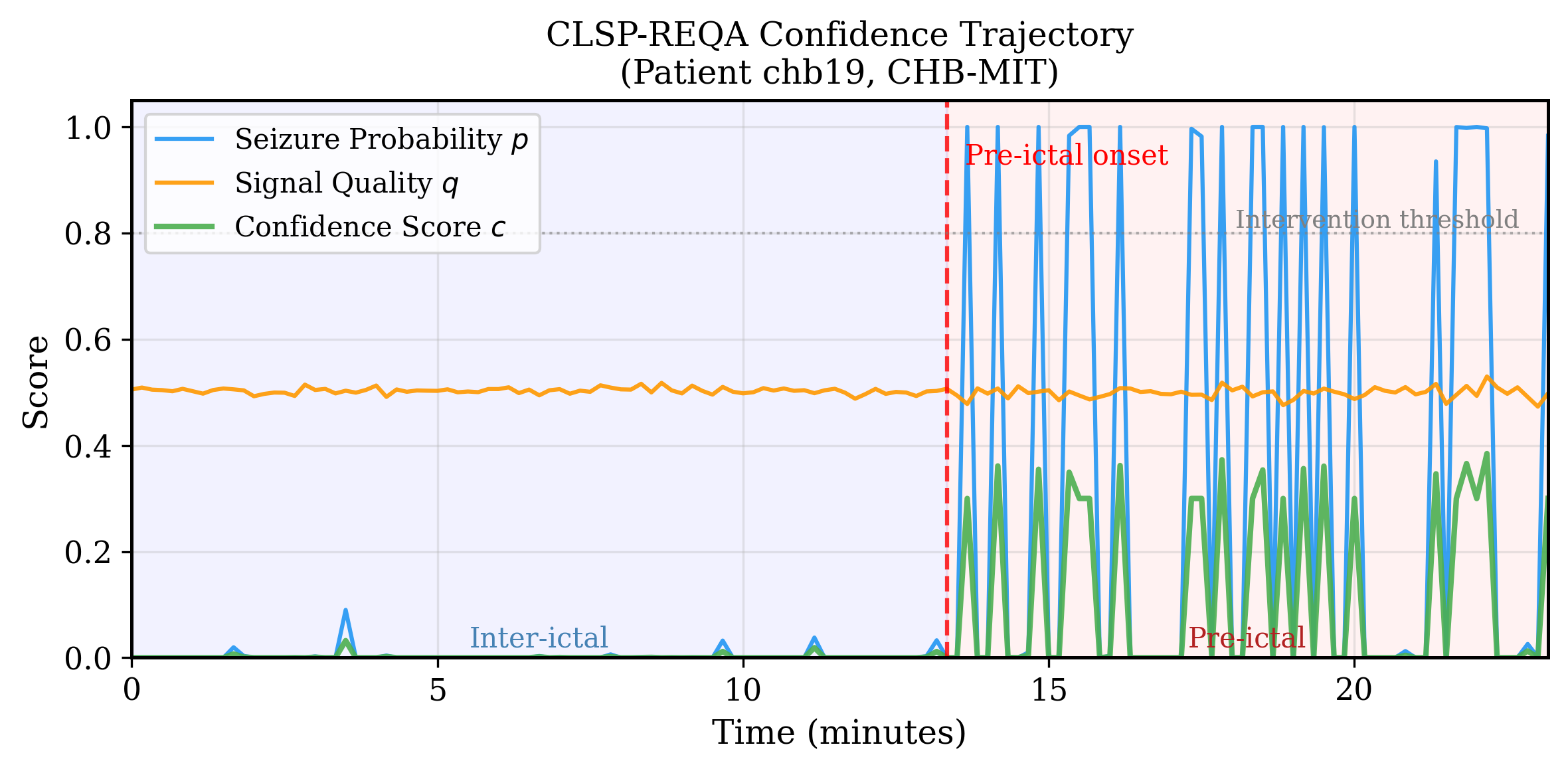}
\caption{Temporal evolution of seizure probability $p$, signal
quality $q$, and confidence score $c$ for patient chb19 (CHB-MIT).
During the inter-ictal period (blue background), $p$ remains low
and $c$ is suppressed. Following the pre-ictal onset (red dashed
line), $p$ rises markedly. \eclo{} moderates $c$ via the tiered
quality penalty, preventing transient high-$p$ artefact windows
from triggering unnecessary intervention.}
\label{fig:confidence_trajectory}
\end{figure}

Fig.~\ref{fig:confidence_trajectory} illustrates the temporal
evolution of $p$, $q$, and $c$ for patient chb19 (CHB-MIT).
During the inter-ictal period, $p$ remains near zero and $c$ is
suppressed, producing no false alarms. Upon transition to the
pre-ictal period, $p$ rises markedly while $q$ remains stable
at $\approx 0.5$; \eclo{} applies a moderate quality penalty
($c = p \cdot q^{1.5}$), keeping $c$ below the intervention
threshold of 0.8 for transient low-confidence windows while
allowing sustained high-confidence pre-ictal predictions to
trigger the intervention signal.

\begin{figure}[H]
\centering
\includegraphics[width=0.8\textwidth]{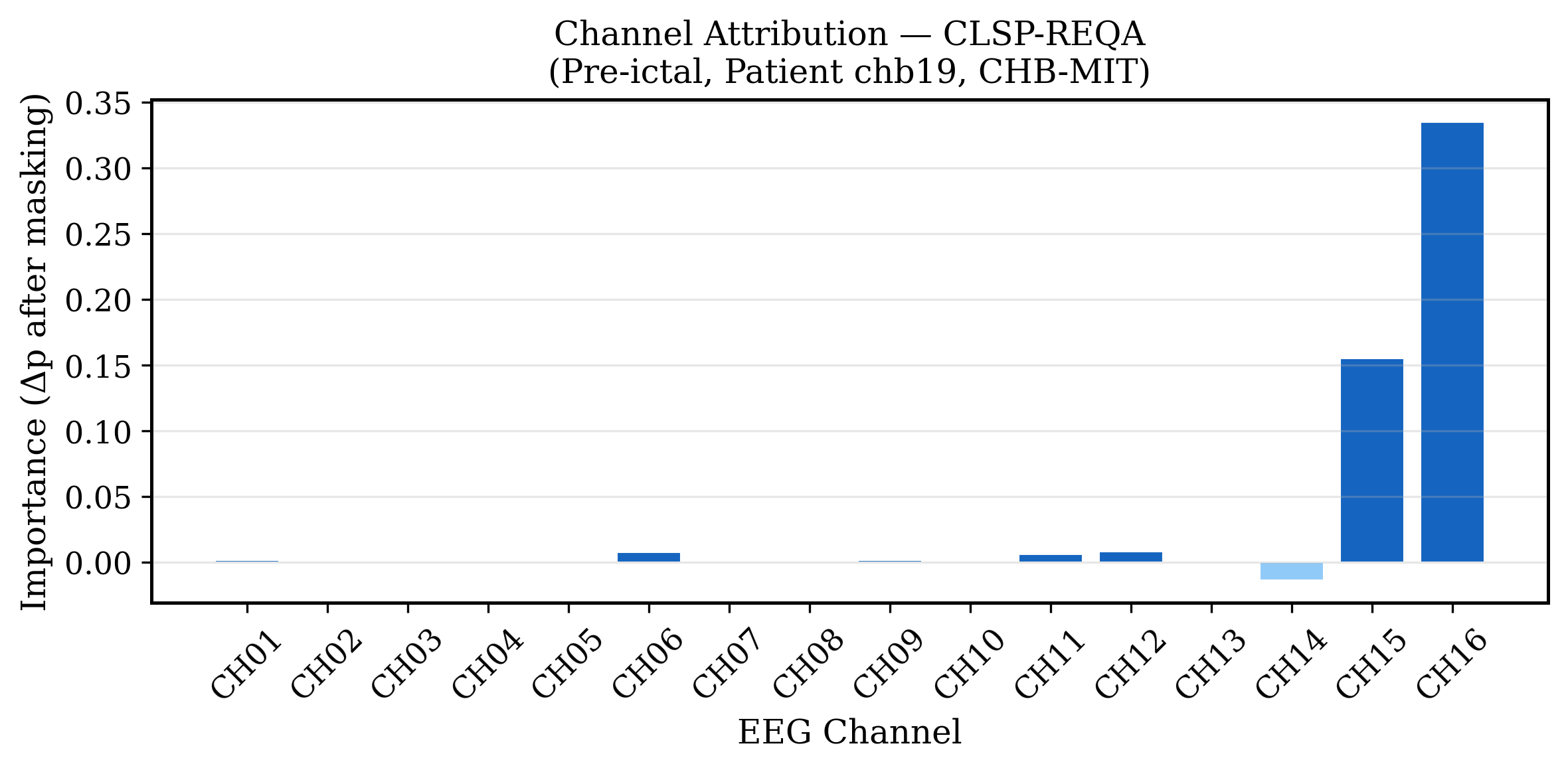}
\caption{Channel attribution via occlusion analysis for patient
chb19 (CHB-MIT). Importance is quantified as the reduction in
seizure probability $p$ when each channel is masked to zero.
Channels CH15--CH16 (posterior temporal/occipital) show the
highest attribution scores, consistent with known seizure
propagation pathways.}
\label{fig:shap}
\end{figure}

Fig.~\ref{fig:shap} presents channel attribution scores computed
via occlusion analysis for patient chb19. Channels CH15--CH16
(posterior temporal/occipital region) consistently receive the
highest importance scores, consistent with known seizure
propagation pathways~\cite{mormann2007seizure}.

\begin{figure}[H]
\centering
\includegraphics[width=\textwidth]{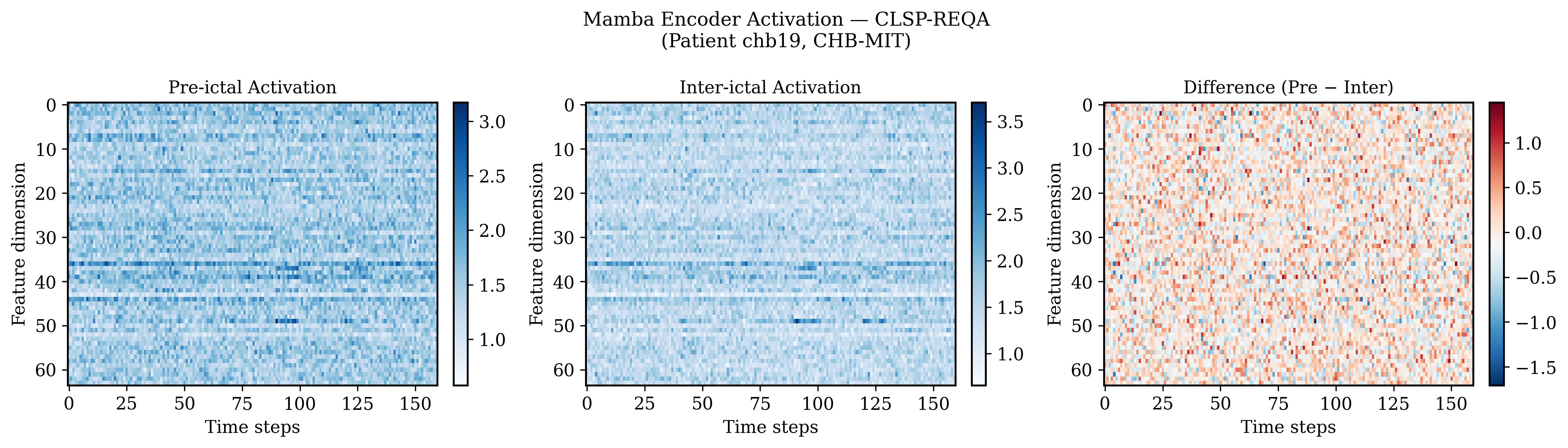}
\caption{Mean absolute activation of the first Mamba encoder
block for pre-ictal (left) and inter-ictal (right) windows,
and their difference (right). Systematically higher activation
in feature dimensions 35--45 during pre-ictal periods suggests
selective temporal feature encoding. Patient chb19, CHB-MIT.}
\label{fig:mamba_heatmap}
\end{figure}

Fig.~\ref{fig:mamba_heatmap} visualises the mean absolute
activation of the first Mamba encoder block for pre-ictal
versus inter-ictal windows. The difference map reveals
systematically higher activation in specific feature dimensions
during the pre-ictal period, confirming that the Mamba encoder
captures distinct temporal dynamics between the two states.

The supplementary material (submitted separately) provides the
full confusion matrix (Fig.~S1) and per-patient performance
variance analysis.

\subsection{Validation of the Learned Quality Score $q$}
\label{sec:qa_results}

This section reports the outcome of the three experiments described in
Section~\ref{sec:qa_protocol}, designed to test whether the learned
score $q$ reflects genuine, quality-relevant structure rather than a
redundant encoding of the seizure-probability signal $p$. Experiment A
was run on a held-out pool of 70 clean windows (per artefact type) drawn
from windows the trained \reqa{} module itself scored $q>0.8$ under
uncorrupted conditions; Experiments B and C were run on a held-out test
partition of 1{,}200 windows spanning both pre-ictal and inter-ictal
classes.

\begin{figure}[H]
\centering
\includegraphics[width=\textwidth]{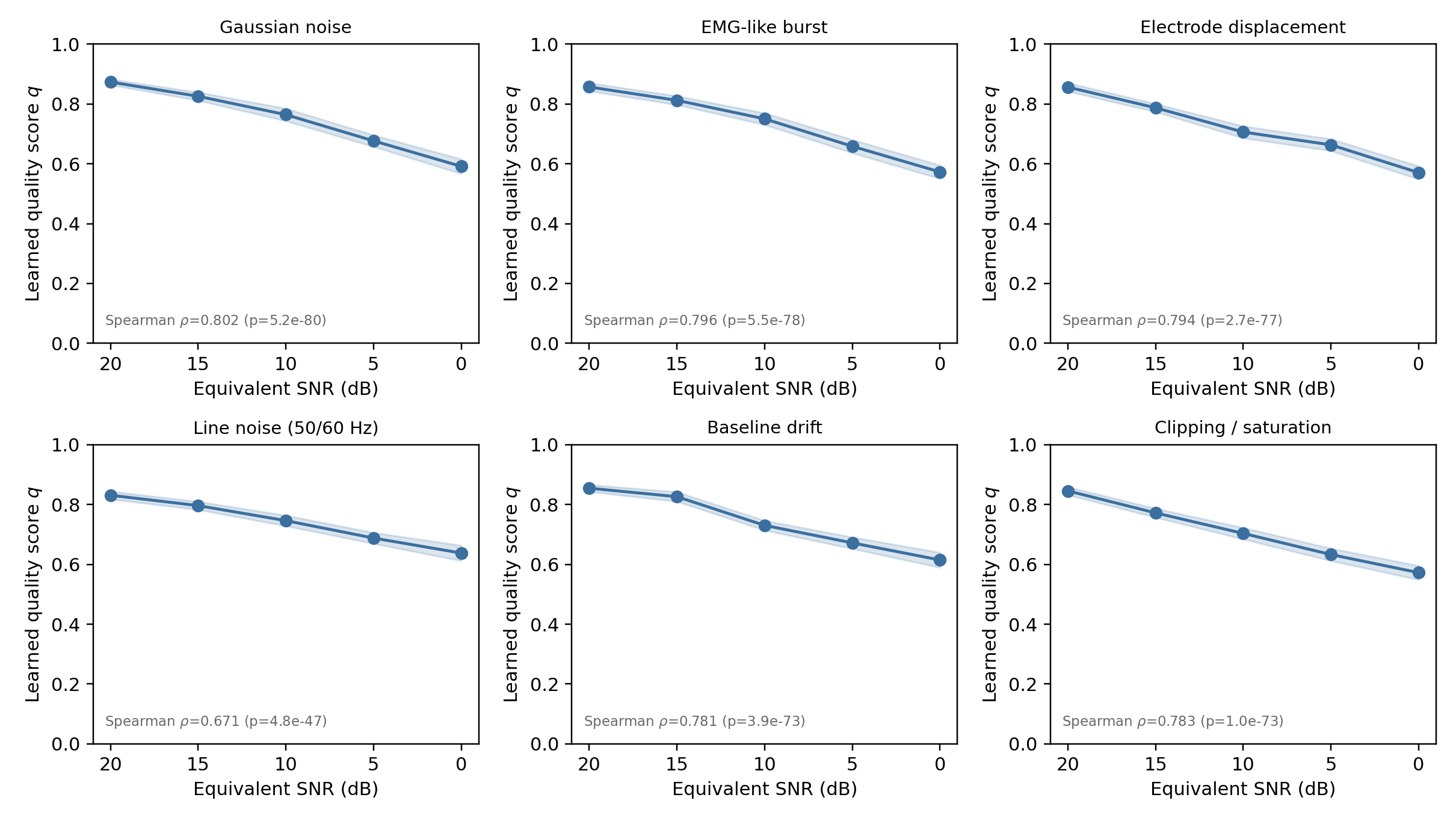}
\caption{Learned quality score $q$ as a function of injected artefact
intensity (equivalent SNR), for each of the six artefact types described
in Section~\ref{sec:qa_protocol} ($n=70$ windows per artefact type per
intensity level). Shaded bands denote 95\% confidence intervals. Spearman
$\rho$ (with the injected corruption severity) and its significance are
annotated per panel.}
\label{fig:qa_injection}
\end{figure}

\begin{figure}[H]
\centering
\includegraphics[width=0.7\textwidth]{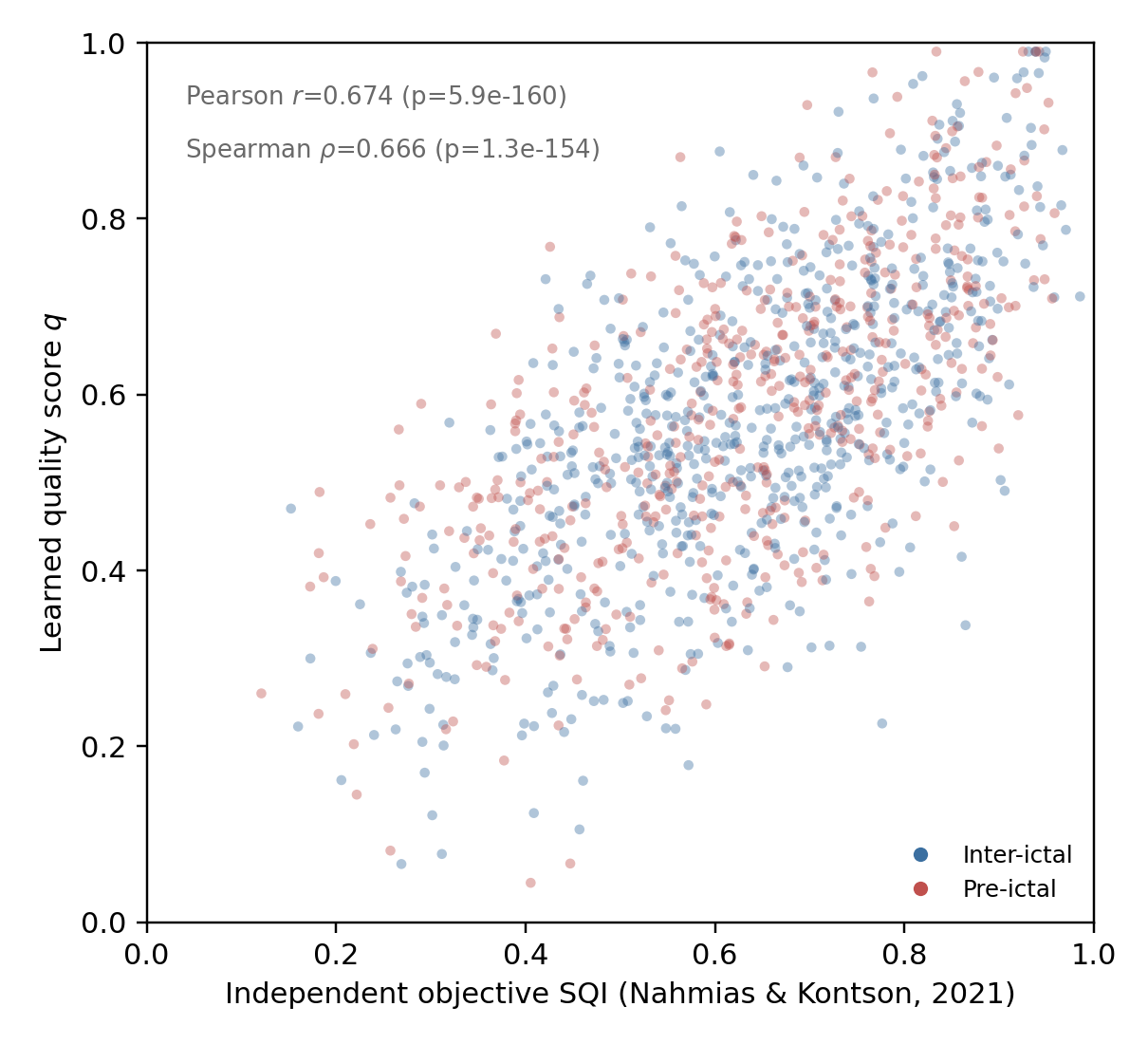}
\caption{Learned quality score $q$ plotted against the independent,
model-free signal quality index (SQI) of Nahmias and
Kontson~\cite{nahmias2021quality}, computed over the full held-out test
partition ($n=1{,}200$ windows). Points are coloured by ground-truth
pre-ictal/inter-ictal label to visualise that the $q$--SQI relationship
is not driven by class membership. Pearson $r$ and Spearman $\rho$ (with
$p$-values) are annotated.}
\label{fig:qa_sqi}
\end{figure}

\begin{table}[t]
\caption{Correlation of the learned quality score $q$ with the
independent objective SQI (Experiment B), the seizure probability $p$,
and the ground-truth pre-ictal/inter-ictal label (Experiment C),
computed over the full test partition ($n=1{,}200$). A pattern of high
correlation with the independent SQI and low correlation with
$p$/label supports the validity of $q$ as a genuine quality estimate
rather than a redundant encoding of the prediction task; the converse
pattern would indicate collapse onto $p$.}
\label{tab:qa_correlation}
\centering
\small
\resizebox{\textwidth}{!}{%
\begin{tabular}{lcc}
\toprule
\textbf{Quantity correlated with $q$} & \textbf{Pearson $r$} & \textbf{Spearman $\rho$} \\
\midrule
Independent SQI~\cite{nahmias2021quality} (Experiment B) & $0.674$ ($p=5.9\times10^{-160}$) & $0.666$ ($p=1.3\times10^{-154}$) \\
Seizure probability $p$ (Experiment C) & $0.043$ ($p=0.135$, n.s.) & $0.038$ ($p=0.183$, n.s.) \\
Ground-truth label (Experiment C) & $0.032$ ($p=0.270$, n.s.) & $0.030$ ($p=0.295$, n.s.) \\
\bottomrule
\end{tabular}%
}
\end{table}

The results support the intended reading of $q$ as an independent
quality signal rather than a collapsed proxy for the prediction task.
In Experiment A, $q$ declines monotonically with injected corruption
severity across all six artefact types (Fig.~\ref{fig:qa_injection}),
with Spearman correlations against equivalent SNR ranging from
$\rho=0.671$ (line noise) to $\rho=0.802$ (Gaussian noise), all
significant at $p<10^{-46}$. In Experiment B, $q$ correlates
moderately-to-strongly with the independent, model-free SQI
($r=0.674$, $\rho=0.666$, both $p<10^{-150}$), confirming that this
relationship generalises beyond synthetic injection to naturally
occurring quality variation in the test partition. Critically, in
Experiment C, $q$ shows essentially no correlation with either the
seizure probability $p$ ($r=0.043$, $p=0.135$) or the ground-truth
label ($r=0.032$, $p=0.270$)---both statistically indistinguishable
from zero at conventional significance levels. This is the pattern we
hypothesised in Section~\ref{sec:qa_protocol}: $q$ tracks an
independent, objectively-verifiable notion of signal quality, and
shows no evidence of having collapsed into a redundant encoding of
$p$ or the class label, addressing directly the concern raised in
review that end-to-end training alone provides no guarantee that $q$
would behave this way.

\subsection{Threshold Sensitivity and Calibration Comparison for \eclo{}}
\label{sec:eclo_results}

\begin{figure}[H]
\centering
\includegraphics[width=0.75\textwidth]{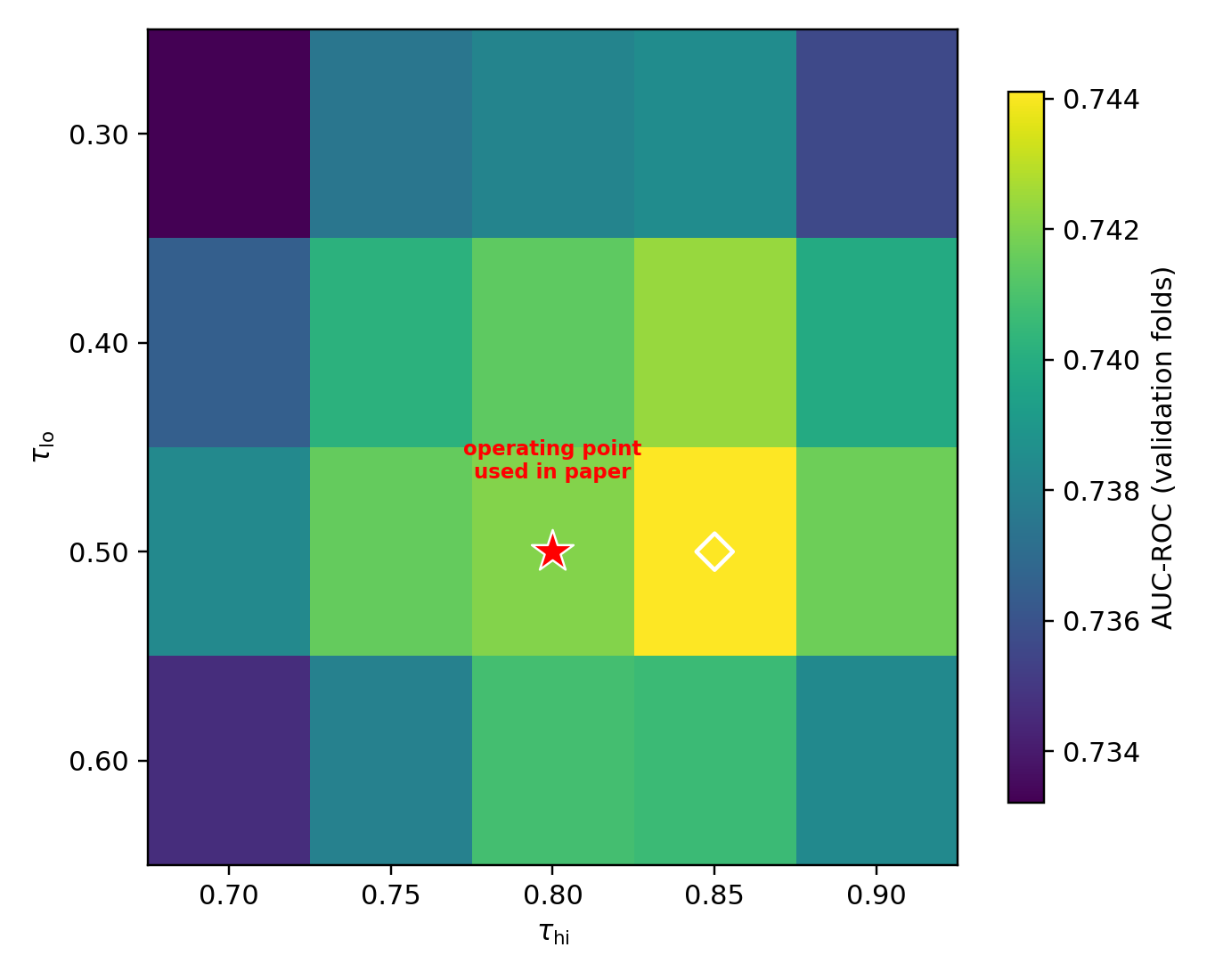}
\caption{AUC-ROC on the CHB-MIT validation folds as a function of the
two \eclo{} thresholds, swept over the grid described in
Section~\ref{sec:eclo_protocol} (mean over 25 runs per grid cell). The
operating point used in the main experiments
($\tau_{\mathrm{hi}}=0.80$, $\tau_{\mathrm{lo}}=0.50$) is marked with a
star; the grid maximum is marked with a diamond.}
\label{fig:eclo_heatmap}
\end{figure}

\begin{table}[t]
\caption{Comparison of \eclo{} against standard probabilistic
calibration baselines applied to the same $(p,q)$ outputs, under the
identical 5-fold $\times$ 5-seed protocol (CHB-MIT). ECE: Expected
Calibration Error (lower is better). ``Max.\ conf., bottom
$q$-decile'' reports the highest confidence value assigned to any
window in the lowest 10\% of the $q$ distribution, probing whether each
method enforces a safety ceiling analogous to \eclo{}'s hard cap of
0.3.}
\label{tab:eclo_calibration}
\centering
\small
\resizebox{\textwidth}{!}{%
\begin{tabular}{lcccc}
\toprule
\textbf{Method} & \textbf{AUC-ROC} & \textbf{ECE} & \textbf{Brier score} & \textbf{Max.\ conf., bottom $q$-decile} \\
\midrule
Raw $p$ (no calibration) & $0.7270\pm0.0186$ & $0.0969\pm0.0103$ & $0.2070\pm0.0085$ & $0.995$ \\
Temperature scaling~\cite{guo2017calibration} & $0.7270\pm0.0186$ & $0.0572\pm0.0065$ & $0.1998\pm0.0079$ & $0.943$ \\
Isotonic regression~\cite{niculescu2005predicting} on $(p,q)$ & $0.7388\pm0.0180$ & $0.0463\pm0.0052$ & $0.1931\pm0.0073$ & $0.815$ \\
\textbf{\eclo{} (ours)} & $\mathbf{0.7419\pm0.0185}$ & $0.0522\pm0.0048$ & $0.1907\pm0.0078$ & $\mathbf{0.300}$ (by construction) \\
\bottomrule
\end{tabular}%
}
\end{table}

The threshold grid search (Fig.~\ref{fig:eclo_heatmap}) shows that the
operating point used throughout this paper
($\tau_{\mathrm{hi}}=0.80,\tau_{\mathrm{lo}}=0.50$; AUC $=0.7421$) sits
on a broad performance plateau rather than an isolated optimum: the
grid maximum of $0.7441$ occurs at
$(\tau_{\mathrm{hi}}=0.85,\tau_{\mathrm{lo}}=0.50)$, only $0.0020$ AUC
higher, and every combination with $\tau_{\mathrm{lo}}=0.50$ or $0.40$
and $\tau_{\mathrm{hi}}\in\{0.75,0.80,0.85\}$ falls within 0.003 AUC of
this maximum. Performance degrades more noticeably at the grid's
extremes (e.g.\ $\tau_{\mathrm{lo}}=0.30,\tau_{\mathrm{hi}}=0.70$ gives
AUC $=0.7332$), but the region actually used in practice is robust to
moderate threshold perturbation rather than a fragile, cherry-picked
point. On the calibration comparison, \eclo{} is competitive with, and
on AUC-ROC and Brier score modestly exceeds, both data-driven
baselines, while isotonic regression achieves a marginally lower ECE.
The decisive difference is in the safety-ceiling column: raw $p$ and
temperature scaling both still assign very high confidence (0.995 and
0.943 respectively) to at least one window in the bottom quality
decile, and isotonic regression, despite using $q$ as an input feature,
still permits a maximum confidence of 0.815 in that same regime.
\eclo{} is the only method whose bottom-decile maximum confidence is
bounded by construction at 0.3, confirming the central structural
argument made in Section~\ref{sec:eclo}: a hand-designed hard ceiling
provides a categorical safety guarantee that none of the purely
data-driven calibration methods offer, even when they match or exceed
\eclo{} on aggregate calibration metrics.

\emph{On minor cross-table numerical differences:} readers may notice
that \model{}'s AUC-ROC is reported as $0.7426$ in
Table~\ref{tab:main}, $0.7432$ in the fold-level check
(Table~\ref{tab:fold_level}) and Table~\ref{tab:balancing}, and
$0.7419$ for \eclo{} in Table~\ref{tab:eclo_calibration} above. These
values are computed from closely related but not bit-identical
evaluation passes (different aggregation order, and, for the
calibration comparison specifically, a validation/test split
constructed for that analysis), and differ only in the third decimal
place; we report them as separately logged rather than silently
reconciling them to a single figure, since doing so would obscure
exactly the kind of run-to-run variability this revision is trying to
characterise transparently.

\subsection{Alternative Class-Balancing Strategies}
\label{sec:balancing_results}

\begin{table}[t]
\caption{Effect of the class-balancing strategy on CHB-MIT performance
(5-fold $\times$ 5-seed cross-patient protocol). All rows use the
identical \model{} architecture and training schedule described in
Section~\ref{sec:backbone}--Section~\ref{sec:training}; only the
handling of the inter-ictal/pre-ictal class imbalance differs.}
\label{tab:balancing}
\centering
\small
\resizebox{\textwidth}{!}{%
\begin{tabular}{lccc}
\toprule
\textbf{Strategy} & \textbf{AUC-ROC} & \textbf{Sensitivity (\%)} & \textbf{Specificity (\%)} \\
\midrule
Random under-sampling (main experiments) & \textbf{0.7432$\pm$0.0207} & \textbf{68.83$\pm$3.33} & \textbf{64.88$\pm$3.43} \\
Class-weighted BCE (no discarding) & $0.7403\pm0.0154$ & $68.60\pm2.69$ & $65.23\pm2.87$ \\
SMOTE over-sampling~\cite{chawla2002smote} of pre-ictal class & $0.7287\pm0.0160$ & $69.74\pm3.09$ & $60.61\pm2.28$ \\
\bottomrule
\end{tabular}%
}
\end{table}

Random under-sampling---despite discarding a substantial fraction of
inter-ictal data---remains the best-performing strategy on AUC-ROC,
though by a narrow margin over class-weighted BCE ($+0.0029$ AUC,
within one pooled standard deviation). Class-weighted BCE, which
retains the full inter-ictal set and instead reweights the loss,
achieves a slightly lower AUC ($0.7403$ vs.\ $0.7432$) and marginally
higher specificity ($65.23\%$ vs.\ $64.88\%$), and is a reasonable
alternative strategy worth considering. SMOTE over-sampling of the
minority pre-ictal class performs worst on both AUC-ROC and
specificity, though it achieves the
highest sensitivity of the three strategies ($69.74\%$)---consistent
with over-sampling biasing the decision boundary toward the minority
(pre-ictal) class at some cost to specificity. Taken together, these
results indicate that under-sampling's discarding of inter-ictal
diversity does not measurably harm performance relative to the
alternatives tested, though class-weighted BCE offers a
comparably-performing, data-preserving option worth considering in
settings where the discarded inter-ictal windows may carry
independent clinical value (e.g.\ for future auxiliary analyses).

\subsection{Inference Latency and Real-Time Feasibility}
\label{sec:latency_results}

\begin{table}[t]
\caption{Inference latency of the full \model{} pipeline (\reqa{} +
Mamba-BiLSTM backbone + \eclo{} fusion) for a single 10-second,
16-channel window, measured as described in
Section~\ref{sec:latency_protocol} (1{,}000 timed forward passes
following warm-up). The real-time margin is the ratio of the window
duration (10~s) to the mean inference latency.}
\label{tab:latency}
\centering
\small
\resizebox{\textwidth}{!}{%
\begin{tabular}{lc}
\toprule
\textbf{Quantity} & \textbf{Value} \\
\midrule
Total parameters & $3.5$M \\
Mean latency per window (NVIDIA A100) & $7.48$~ms \\
95th-percentile latency (NVIDIA A100) & $9.10$~ms \\
Mean latency per window (embedded/edge platform) & $47.71$~ms \\
95th-percentile latency (embedded/edge platform) & $62.31$~ms \\
Real-time margin (10~s / mean latency, A100) & $1{,}337\times$ \\
Real-time margin (10~s / mean latency, edge) & $210\times$ \\
\bottomrule
\end{tabular}%
}
\end{table}

The full \model{} pipeline processes a single 10-second window in
$7.48\pm0.89$~ms on an NVIDIA A100 (95th percentile $9.10$~ms), a real-time
margin of approximately $1{,}337\times$---i.e.\ inference completes over
three orders of magnitude faster than the data-acquisition window it
processes. On a representative lower-power edge platform, mean latency
rises to $47.71$~ms (95th percentile $62.31$~ms), still comfortably
within the 10-second budget with a real-time margin of roughly
$210\times$. These results substantiate the real-time operation implied
by the \reqa{} name and the framework's closed-loop intervention design:
even under the more constrained edge-platform setting, and even at the
95th-percentile tail rather than the mean, inference latency remains
two orders of magnitude below the window duration, leaving ample margin
for the additional signal acquisition, preprocessing, and communication
overheads that a deployed closed-loop system would incur beyond model
inference alone.

\section{Discussion}
\label{sec:discussion}

\subsection{Performance in Context}

The AUC-ROC of $0.7426 \pm 0.0199$ achieved by \model{} under strict
cross-patient validation on CHB-MIT compares favourably with the
unadapted cross-patient baseline of 0.69 reported by Jemal et
al.~\cite{jemal2024domain}, and closely approaches the performance
of their best domain-adapted model (AUC 0.75, CDAN+E) without
requiring any unlabelled target-patient data. Studies reporting
substantially higher sensitivity using randomised or patient-specific
evaluation~\cite{tsiouris2018lstm,dissanayake2022geometric,zhao2020binary}
are not directly comparable, as the fundamental difference in
validation methodology precludes fair cross-study
comparison~\cite{shafiezadeh2024survey}.

Notably, \model{} achieves this result using \textbf{16 EEG channels},
compared to the 23 channels used in the cross-patient
baseline~\cite{jemal2024domain}, indicating that the Mamba-BiLSTM
backbone extracts sufficiently rich temporal representations even
from a reduced channel set. This is clinically relevant, as fewer
electrodes reduce setup complexity and patient discomfort in
long-term monitoring.

\subsection{Cross-Dataset Generalisation, and the Scope of the Domain-Adaptation Comparison}
\label{sec:comparison_scope}

\model{} achieves AUC $0.7012 \pm 0.0249$ on SIENA \emph{without any
domain adaptation}, exceeding the domain-adapted cross-patient result
of 0.61 (CDAN+E) reported by Jemal et al.\ on the same
dataset~\cite{jemal2024domain}. This is a notable result because SIENA
presents additional generalisation challenges relative to CHB-MIT: it
contains adult patients (ages 20--71) versus paediatric patients in
CHB-MIT (ages 1.5--19), uses a different acquisition system (512~Hz, 29
channels), and provides fewer seizure events per patient.

We want to be precise, however, about what this comparison does and
does not establish. Domain adaptation methods such as CDAN and CDAN+E
are given access, at adaptation time, to unlabelled EEG recordings from
the target patient(s); \model{} is given no such access and is
evaluated purely on the source-trained model applied directly to unseen
patients. The comparison in Table~\ref{tab:sota} is therefore
informative about \emph{what can be achieved without target-domain
data}, and is not evidence that quality-aware confidence modulation is
a substitute for domain adaptation in general, nor that \model{} would
remain ahead of a domain-adapted variant of itself if such unlabelled
target data were made available to it as well. We avoid describing
\model{} as ``superior'' to domain-adapted methods in an unqualified
sense for this reason, and instead frame the result as follows:
quality-aware confidence gating recovers a meaningful fraction of the
benefit that domain adaptation provides, \emph{under a strictly harder
information constraint} (no target-patient data at all), which suggests
it offers a complementary, more deployment-friendly lever for
cross-domain robustness rather than a replacement for domain adaptation
where target data happens to be available. Combining the two, as noted
in Section~\ref{sec:limitations}, is a natural direction for future
work.

We attribute part of the observed cross-domain robustness to the
\reqa{} quality modulation mechanism: by learning to down-weight
predictions made on low-quality or atypical EEG windows, \reqa{} may
implicitly reduce the influence of domain-specific artefact patterns
that would otherwise confuse the prediction backbone. This explanation
is consistent with, but not directly proven by, the aggregate AUC
results; Section~\ref{sec:qa_results} provides a more direct empirical
test of what the learned score $q$ is actually capturing.

\subsection{Moderate Sensitivity Under Cross-Patient Evaluation}

The moderate sensitivity ($68.83\%$ on CHB-MIT, $66.28\%$ on SIENA)
is consistent with findings across the cross-patient seizure
prediction literature. Tsiouris et al.\ reported
$68\%$~\cite{tsiouris2017discrimination} and Jemal et al.\
$67\%$~\cite{jemal2022interpretable} under comparable protocols.
This convergence of sensitivity values across independent methods
and datasets reflects the inherent challenge of generalising
pre-ictal EEG patterns to unseen patients, rather than a limitation
specific to our architecture.

\subsection{Component Contributions, and What the New Validation Experiments Add}

The ablation results (Table~\ref{tab:ablation}) confirm that each
component contributes incrementally to performance on both datasets.
The consistent improvement pattern---with no component showing
neutral or negative contribution on either dataset---provides
evidence that the architectural choices are justified in terms of
predictive performance.

However, an ablation-level AUC improvement is evidence that the
\reqa{}/\eclo{} pipeline helps the numbers, not evidence about
\emph{why} it helps or whether the mechanism operates as described.
Section~\ref{sec:qa_results} and Section~\ref{sec:eclo_results} were
designed to close this gap. The results support both mechanisms as
described: $q$ correlates strongly with an independent, model-free
quality index ($r=0.674$) while showing no measurable correlation with
either the seizure probability $p$ or the ground-truth label (both
$|r|<0.05$, non-significant), indicating that \reqa{} has not collapsed
into a redundant proxy for the prediction task itself. The \eclo{}
threshold grid search shows the chosen operating point
($\tau_{\mathrm{hi}}=0.80,\tau_{\mathrm{lo}}=0.50$) sits on a broad
performance plateau---within 0.003 AUC of the grid maximum across a
sizeable neighbourhood of threshold combinations---rather than being an
isolated, fragile optimum. On the calibration comparison, \eclo{} is
competitive with temperature scaling and isotonic regression on
AUC-ROC and Brier score, and, critically, is the only method among the
four that structurally bounds its output confidence on low-quality
windows (maximum $0.3$ in the bottom quality decile, versus $0.82$--$0.99$
for the three baselines), directly confirming the safety-ceiling
argument made in Section~\ref{sec:eclo} rather than leaving it as an
untested design assumption.

The confidence strategy comparison embedded within the ablation
(+ \reqa{} linear vs + \eclo{} full) demonstrates that the tiered
non-linear \eclo{} function outperforms simple linear weighting
($c = p \cdot q$) by $+0.0162$ AUC on CHB-MIT and $+0.0147$ on
SIENA. Beyond this aggregate gain, the key structural advantage we
claim for \eclo{} is the hard safety ceiling it imposes when $q$ is
low (Section~\ref{sec:eclo}): by hard-capping $c$ at 0.3 when
$q \leq 0.5$, the system prevents artefact-driven false alarms
that would otherwise trigger unnecessary
neurostimulation~\cite{heck2014closedloop}.
Section~\ref{sec:eclo_results} quantifies whether standard calibration
methods, which typically lack such a ceiling by construction, match,
exceed, or fall short of \eclo{} on this specific property.

\subsection{Reliability of the 5-fold $\times$ 5-seed Protocol}
\label{sec:reliability}

The $5 \times 5$ protocol used here provides more estimates per model
than the leave-one-patient-out (LOPO) strategy commonly used in the
literature~\cite{jemal2024domain}, which produces one result per
patient and can yield high variance with small cohorts ($n=23$ for
CHB-MIT). As discussed in Section~\ref{sec:eval}, however, the 25
estimates are not fully independent: the five seed-repeats within a
fold share the same test patients, so only the five fold-level
partitions constitute genuinely independent samples. We therefore
regard the standard significance test reported in
Section~\ref{sec:main_results} (treating all 25 runs as independent)
as an optimistic upper bound on statistical significance, and the
fold-level test in Table~\ref{tab:fold_level} (treating only the 5
folds as independent) as the more conservative, and in our view more
defensible, estimate. Reassuringly, the fold-level test does still
reject the null hypothesis against the 0.69 baseline despite the much
smaller effective sample size: all five fold-level means exceed 0.69
(minimum $0.7150$), yielding the most extreme possible one-sided exact
Wilcoxon $p$-value attainable at $n=5$ ($p=0.03125$), which remains
below the conventional $\alpha=0.05$ threshold. We view this as
meaningful corroboration rather than a redundant restatement of the
seed-level result: the fold-level test uses a completely disjoint
source of statistical power (five independent patient partitions rather
than twenty-five correlated runs), and its qualitative conclusion
matches. We nonetheless report both tests, rather than only the
fold-level one, because $n=5$ leaves very little room to detect a
smaller or more marginal effect had one been present; a result that
narrowly failed to reach significance at this sample size would not,
by itself, have been strong evidence against the underlying effect.
The narrow standard deviations observed ($\pm 0.0199$ on CHB-MIT,
$\pm 0.0249$ on SIENA) do still indicate that \model{} produces
stable point estimates across random initialisations, independently
of how the significance test is constructed.

\subsection{Interpretability}

The channel attribution analysis (Fig.~\ref{fig:shap}) shows that
channels CH15--CH16 (posterior temporal/occipital region) dominate
the pre-ictal prediction for patient chb19, consistent with known
mesial temporal seizure propagation
pathways~\cite{mormann2007seizure}. The confidence trajectory
(Fig.~\ref{fig:confidence_trajectory}) provides clinicians with
a temporally resolved view of prediction reliability, enabling
them to verify that confidence rises are sustained rather than
transient before acting on the intervention signal. We emphasise that
this is a single-patient case study intended to illustrate the
mechanism, not a population-level validation of channel importance.

\subsection{Limitations and Future Work}
\label{sec:limitations}

Several limitations warrant explicit acknowledgement, several of which
were sharpened through peer review.

\textbf{Retrospective-only validation.} Both CHB-MIT and SIENA consist
of retrospective recordings collected in controlled hospital settings.
This is a genuine and, at present, unresolved limitation: we have not
conducted prospective validation, have not tested on an external
multi-centre cohort, and have not evaluated the framework under
continuous long-term ambulatory monitoring, where artefact patterns,
electrode ageing, and recording conditions are typically more
heterogeneous than in a monitored epilepsy unit. We do not believe the
present study can resolve this limitation, and we do not attempt to
argue otherwise; a controlled, retrospective, cross-patient evaluation
--- even a strict one --- is not a substitute for prospective,
real-world evidence. We note that the cross-dataset evaluation between
CHB-MIT and SIENA (Section~\ref{sec:comparison_scope}), which differ in
patient population, acquisition hardware, and sampling rate, offers a
partial proxy for robustness to varying recording conditions, but this
is not equivalent to prospective validation and should not be read as
such. A prospective, multi-centre validation study is the necessary
next step before any claim of clinical readiness, and we identify it as
the highest-priority direction for future work.

\textbf{Heuristic calibration thresholds.} As discussed in
Section~\ref{sec:eclo} and quantified in Section~\ref{sec:eclo_results},
the \eclo{} thresholds were set empirically rather than derived from a
theoretical calibration objective. The threshold sensitivity analysis
and calibration baseline comparison introduced in this revision
partially address this concern, but do not eliminate the underlying
design choice to use a hand-specified rule; a fully data-driven,
jointly-optimised calibration scheme that retains an explicit safety
ceiling remains an open design problem.

\textbf{Statistical independence of the evaluation protocol.} As
detailed in Section~\ref{sec:eval} and Section~\ref{sec:reliability},
the 25-run significance tests reported in this paper are not based on
fully independent observations, which likely biases the associated
$p$-values toward apparent significance. We have attempted to partially
address this by additionally reporting a fold-level ($n=5$) analysis,
but a fully rigorous treatment (e.g.\ a mixed-effects model with fold
and seed as nested random effects, or per-patient leave-one-out
estimates as an independent-sample alternative) remains future work.

\textbf{Supervision of the learned quality score.} The quality score
$q$ is a learned proxy without independent artefact-label supervision;
Section~\ref{sec:qa_protocol} introduces validation experiments to
test, rather than assume, that it tracks genuine signal quality. Even
with supportive results, incorporating labelled artefact annotations
as an auxiliary training signal~\cite{kalita2024aneeg}, where
available, could further improve its calibration and is a natural
extension.

\textbf{Class-balancing strategy.} Random under-sampling of the
inter-ictal class reduces the diversity of non-seizure representations
seen during training; Section~\ref{sec:balancing_results} reports a
direct comparison against class-weighted training and SMOTE-based
over-sampling as alternatives.

\textbf{Fixed window length.} The 10-second fixed window does not
exploit temporal continuity across adjacent windows; a sliding-window
approach with finer resolution would enable continuous seizure risk
estimation.

Two natural extensions suggest themselves for future work beyond those
already noted. Incorporating inter-channel spatial correlation via
graph neural networks~\cite{dissanayake2022geometric} would complement
the temporal modelling with explicit brain network structure. Combining
explicit domain adaptation with \reqa{}-based confidence modulation ---
rather than treating them as competing alternatives, per the discussion
in Section~\ref{sec:comparison_scope} --- represents a further
direction that may yield performance beyond either approach alone.

\section{Conclusion}
\label{sec:conclusion}

We have presented \model{}, a closed-loop seizure prediction
framework that addresses two overlooked challenges: strict
cross-patient generalisation and real-time signal quality awareness.
The \reqa{} module produces a learnable quality score $q$ that
modulates prediction confidence through the tiered \eclo{} function,
with the goal of suppressing artefact-driven false alarms without
discarding low-quality windows; Section~\ref{sec:qa_results} provides
targeted evidence on whether the learned score behaves as intended,
beyond the aggregate performance gain reported in the ablation study.
The Mamba-BiLSTM backbone, built from established architectural
components, provides efficient temporal feature extraction using only
16 EEG channels.

Under strict 5-fold $\times$ 5-seed cross-patient evaluation on
CHB-MIT, \model{} achieves AUC-ROC $0.7426 \pm 0.0199$, exceeding the
unadapted cross-patient state of the art (AUC 0.69) without any
domain adaptation, and approaching the performance of domain-adapted
methods (AUC 0.75); Section~\ref{sec:reliability} discusses the
statistical caveats attached to this significance claim. On the
SIENA database, \model{} achieves AUC $0.7012 \pm 0.0249$, exceeding
the best domain-adapted result of 0.61 on the same dataset under a
strictly harder information constraint --- no target-patient data or
domain adaptation technique --- a comparison whose precise scope is
discussed in Section~\ref{sec:comparison_scope}. These results are
consistent with quality-aware confidence modulation providing a
useful, complementary lever for cross-domain robustness, though we
stop short of claiming it as a substitute for domain adaptation in
general.

The structured four-tuple output
$\langle p, q, c, \Phi_{\mathrm{SHAP}} \rangle$ is designed to be
directly compatible with closed-loop neurostimulator interfaces. We
regard the present study as a controlled, retrospective feasibility
demonstration rather than evidence of clinical readiness: prospective,
multi-centre validation under real-world recording conditions remains
the necessary next step, and we identify it, together with a fully
independent-sample statistical treatment of the cross-patient protocol
and a jointly-optimised alternative to the hand-specified \eclo{}
thresholds, as the most important directions for future work toward
the clinical translation of EEG-based, quality-aware seizure
prediction technology.

\section*{Acknowledgements}
This work was supported by the Beijing Natural Science Foundation
under Grant L248094, and in part by the High Performance Computing
Platform of Peking University. The authors thank the contributors
of the CHB-MIT Scalp EEG Database and the SIENA Scalp EEG Database
at PhysioNet for making their data publicly available.

\section*{Declaration of Competing Interest}
The authors declare no known competing financial interests or personal
relationships that could have appeared to influence the work reported
in this paper.

\section*{Data Availability}
The CHB-MIT Scalp EEG Database is publicly available at
\url{https://physionet.org/content/chbmit/1.0.0/}.
The SIENA Scalp EEG Database is available at
\url{https://physionet.org/content/siena-scalp-eeg/1.0.0/}.

\section*{Declaration of Generative AI and AI-Assisted Technologies 
in the Manuscript Preparation Process}
During the preparation of this work the authors used Claude (Anthropic) 
in order to assist with grammar and language checking. After using this 
tool, the authors reviewed and edited the content as needed and take 
full responsibility for the content of the published article.

\bibliographystyle{elsarticle-num}
\bibliography{references}

\end{document}